\documentclass[graybox]{svmult}

\usepackage{mathptmx}       % selects Times Roman as basic font
\usepackage{helvet}         % selects Helvetica as sans-serif font
\usepackage{courier}        % selects Courier as typewriter font
\usepackage{type1cm}        % activate if the above 3 fonts are not available on your system
\usepackage{amssymb}
\usepackage{amsmath}
\usepackage{amsfonts}
\usepackage{pictexwd,dcpic}
\usepackage{makeidx}         % allows index generation
\usepackage{graphicx}        % standard LaTeX graphics tool
\usepackage{multicol}        % used for the two-column index
\usepackage[bottom]{footmisc}% places footnotes at page bottom
\usepackage{tensor}
\usepackage{eucal,mathrsfs}

\makeindex             % used for the subject index
\def\a{\alpha}

\def\g{\gamma}

\def\p{{\mathbf{p}}}

\newcommand{\beq}{\begin{equation}}

\newcommand{\beql}[1]{\begin{equation}\label{#1}}
\newcommand{\eeq}{\end{equation}}
\newcommand{\bseq}{\begin{subequations}}
\newcommand{\bseql}[1]{\begin{subequations}\label{#1}}
\newcommand{\eseq}{\end{subequations}}

\newcommand{\ft}{\protect\footnote}

\newcommand{\rt}{\sqrt}
\newcommand{\fr}{\frac}
\newcommand{\tn}{\tensor}

\newcommand{\de}{\delta}

\newcommand{\Dl}{\nabla}

\newcommand{\intdx}[1]{\int\mathrm{d}^{#1}x\,}

\newcommand{\Gae}{G_{\ae}}

\newcommand{\met}{\mathsf{g}}

\newcommand{\ac}{\CMcal{S}}
\newcommand{\acEH}{\ac_\text{\sc eh}}
\newcommand{\lag}{\mathscr{L}}

\newcommand{\meV}{{\rm meV}}
\newcommand{\eV}{{\rm eV}}
\newcommand{\keV}{{\rm keV}}
\newcommand{\MeV}{{\rm MeV}}
\newcommand{\GeV}{{\rm GeV}}
\newcommand{\TeV}{{\rm TeV}}

\newcommand{\kpc}{{\rm kpc}}

\newcommand{\s}{{\rm s}}

\newcommand{\Mpl}{M_{\rm Pl}}

\begin{document}
\title*{Lorentz breaking effective field theory models for matter and gravity: theory and observational constraints}
%\title*{Lorentz symmetry violation in effective field theory and observational constraints}
% your contribution title if the original one is too long\
\titlerunning{Lorentz breaking effective field theory models}
%\and Name of Second Author \at Name, Address of Institute \email{name@email.address}
\author{Stefano Liberati \and David Mattingly}
\institute{Stefano Liberati \at SISSA and INFN, Via Bonomea 265, Trieste, Italy \email{liberati@sissa.it}
\and David Mattingly \at 
%Department of Physics, 
University of New Hampshire, Durham, NH 03824, USA \email{dyo7@unh.edu}}

\maketitle

\abstract{A number of different approaches to quantum gravity are at least partly phenomenologically characterized by their treatment of Lorentz symmetry, in particular whether the symmetry is exact or modified/broken at the smallest scales.  For example, string theory generally preserves Lorentz symmetry while analog gravity and Lifshitz models break it at microscopic scales. In models with broken Lorentz symmetry there are a vast number of constraints on departures from Lorentz invariance that can be established with low energy experiments by employing the techniques of effective field theory in both the matter and gravitational sectors.  We shall review here the low energy effective field theory approach to Lorentz breaking in these sectors and present various constraints provided by available observations.}

\section{Introduction}
\label{sec:1}

Our understanding of the observed laws of Nature is currently based on two different theories: the Standard Model of Particle Physics (SM), and General Relativity (GR).  However, in spite of their phenomenological successes, SM and GR leave many fundamental theoretical questions unanswered. First of all, part of the success of the SM has been the recognition that symmetry breaking is an important part of modern physics and that what appear to be multiple forces at low energies can often be described in a unified manner.  The prime example of this is the Glashow-Salam-Weinberg theory of electroweak interactions.  Since such unification is possible, and since many physicists feel that our understanding of the fundamental laws of Nature is deeper and more accomplished if we are able to reduce the number of degrees of freedom in a theory, much effort has been spent trying to construct unified theories in which not only all the sub-nuclear forces are seen as different aspects of a unique interaction, but also gravity is included in a consistent manner as merely part of the overall structure.  

Another important reason why we seek for a new theory of gravity comes directly from the gravity side. We know that GR fails to be a predictive theory in some regimes. Indeed, many solutions of Einstein's equations are singular in some region, and GR is not able to make any prediction in those regions of spacetime. Moreover, there are honest classical solutions of the Einstein equations that contain closed time-like curves, which would allow traveling back and forth in time with the associated causal paradoxes.  Finally, the problem of black-hole evaporation considered just within the framework of semi-classical gravity clashes with quantum mechanical unitary evolution. 

This long list of puzzles spurred intense research toward a quantum theory of gravity that started almost immediately after Einstein's proposal of GR and which is still one of the most active areas of theoretical physics. The quantum gravity problem is not only conceptually and technially challenging, it has also been an almost metaphysical pursuit for several decades, in that what progress has been made is on the theoretical side and the experimental aspect has been (for good reason) neglected. Indeed, we expect QG effects at experimentally/observationally accessible energies to be extremely small, due to suppression by the Planck scale $M_{\rm pl} \equiv \sqrt{\hbar c/G_{\rm N}}\simeq 1.22\times 10^{19}~\mbox{GeV}/c^{2}$. In this sense it has been considered (and it is still considered by many) that only ultra-high-precision (or Planck scale energy) experiments would be able to test quantum gravity models.

It was however realized (mainly over the course of the past decade) that the situation is not quite as bleak as it appears. In fact, quantum gravitational models beyond GR  have shown that there can be several of what we term low energy ``relic signatures'' of quantum gravitational effects which would lead to deviation from the standard theory predictions (SM plus GR) in specific regimes. Some of these new phenomena, which comprise what is often termed ``QG phenomenology'', include: 
\begin{itemize}
\item Quantum decoherence and state collapse \cite{Mavromatos:2004sz}
\item QG imprint on initial cosmological perturbations \cite{Weinberg:2005vy}
\item Cosmological variation of couplings \cite{Damour:1994zq,Barrow:1997qh}
\item TeV Black Holes, related to extra-dimensions \cite{Bleicher:2001kh}
\item Violation of discrete symmetries \cite{Kostelecky:2003fs}
\item Violation of space-time symmetries \cite{Mattingly:2005re}
\end{itemize}
In this chapter we will focus upon the phenomenology of violations of space-time symmetries, and in particular of Local Lorentz invariance (LLI), a pillar both of quantum field theory as well as GR (LLI is a crucial part of the Einstein Equivalence Principle on which metric theories of gravity are based). 

\section{A brief history of Lorentz breaking}
\label{sec:LVhistory}

Contrary to the common perception, explorations of the possible breakdown of LLI have a long standing history. It is however undeniable that the last twenty years have witnessed a striking acceleration in the development both of theoretical ideas as well as of phenomenological tests previously unimagined. We shall here present an admittedly incomplete review of these developments.

\subsection{Early works}

The possibility that Lorentz invariance violation (LV) could play a role in physics dates back at least sixty years~\cite{DiracAet,Bj,Phillips66,Blokh66,Pavl67,Redei67} and in the seventies and eighties there was already a well established literature investigating the possible phenomenological consequences of LV (see e.g.~\cite{1978NuPhB.141..153N,1980NuPhB.176...61E,Zee:1981sy,Nielsen:1982kx,1983NuPhB.217..125C,1983NuPhB.211..269N}).  

The relative scarcity of these studies in the field was due to the general expectation that new effects would only appear in particle interactions where the particle energies were of order the Planck scale, as only at those energies would the natural QG suppression by powers of the Planck scale be overcome.  However, it was only in the nineties that it was clearly realized that there are special situations in which new effects, even if highly suppressed, can have observational consequences. These situations were termed ``Windows on Quantum Gravity''.

\subsection{The dawn of Quantum Gravity Phenomenology}

At first glance, it appears hopeless to search for effects suppressed by the Planck scale. Even the most energetic particles ever detected (Ultra High Energy Cosmic Rays, see, e.g.,~\cite{Roth:2007in,Abbasi:2007sv}) have $E \lesssim 10^{11}$ GeV $\sim 10^{-8} M_{\mathrm pl}$. 
However, tiny corrections can be magnified into an observational effect if the physics in question involves not just the Planck scale and the energy scale of the particle, but also another scale such as the mass of a light particle or a cosmological travel time.  In these situations observables can be constructed that leverage the scales against one another to create an actually measurable effect at achievable energies (i.e. anything from the energies in tabletop optical experiments to cosmic rays. See, e.g.,~\cite{Mattingly:2005re, Bluhm:2005uj} for an extensive review).

A partial list of these {\em windows on QG} includes:
\begin{itemize}
\item sidereal signal variations as a laboratory apparatus such as an optical cavity moves
  with respect to a preferred frame or direction
\item cumulative effects: long baseline dispersion and vacuum birefringence (e.g.~of signals from gamma ray bursts, active galactic nuclei, pulsars)
\item anomalous (normally forbidden) threshold reactions allowed by LV terms (e.g.~photon decay, vacuum \v{C}erenkov effect) 
\item shifting of existing threshold reactions (e.g.~photon annihilation from Blazars, ultra high energy protons pion production)
\item LV induced decays not characterized by a threshold (e.g.~decay of a particle from one helicity to the other or photon splitting)
\item maximum attainable particle velocities different from c (e.g.~synchrotron peak from supernova remnants)
\item dynamical effects of LV background fields (e.g. gravitational coupling and additional wave modes)
\end{itemize}

It is rare one can assign a definitive ``paternity" to a field, and our so called ``Quantum Gravity Phenomenology'' is no exception.  However, among the papers commonly accepted as seminal we can cite  the one by Kosteleck\'{y} and Samuel \cite{KS89} that already in 1989 envisaged, within a string field theory framework, the possibility of non-zero vacuum expectation values (VEV) for some Lorentz breaking operators. This work led later on to the development of a systematic extension of the SM (what was later on called ``minimal standard model extension" (mSME)) incorporating all possible Lorentz breaking, power counting renormalizable operators (i.e. of mass dimension $\leq 4$) by Colladay and Kosteleck\'{y}~\cite{Colladay:1998fq}.  This provided a framework for computing in effective field theory the observable consequences for many experiments and led to much experimental work setting limits on the LV parameters in the Lagrangian (see e.g.~\cite{Kostelecky:2008zz} for a review).
 
Another seminal paper was that of Amelino-Camelia and collaborators~\cite{AmelinoCamelia:1997gz} which highlighted the possibility to cast observational constraints on Planck-suppressed violations of Lorentz invariance in the photon dispersion relation by examining the propagation of light from remote astrophysical sources like gamma ray bursters (GRBs) and active galactic nuclei (AGN). Finally, we also mention the influential papers by Coleman and Glashow \cite{Coleman:1997xq,Coleman:1998en,Coleman:1998ti} which brought the subject of systematic tests of Lorentz violation to the attention of of the broader community of particle physicists. 

Let us stress that this is necessarily an incomplete account of the literature that investigated departures from Special Relativity. Several papers appeared in the same period and some of them anticipated many important results, see e.g.\cite{GonzalezMestres:1996zv,GonzalezMestres:1997if}; unfortunately at the time of their appearance they were hardly noticed (and seen by many as too ``exotic").

In the first decade after 2000 the field reached a concrete maturity and many papers pursued both a systematization of the various frameworks and the available constraints (see e.g.~\cite{Jacobson:2002hd, Mattingly:2002ba, Jacobson:2005bg}). In this sense another crucial contribution was the development of an effective field theory approach also for higher order (mass dimension greater than four), naively non-power counting renormalizable, operators \footnote{Anisotropic scaling \cite{Anselmi:2008ry,Horava:2009uw,Visser:2009ys} techniques were recently recognized to be the most appropriate way of handling higher order operators in Lorentz breaking theories and in this case the highest order operators are indeed crucial in making the theory power counting renormalizable. This is why we shall adopt sometimes the expression ``naively non renormalizable"}. This was first done for rotationally invariant dimension 5 operators in QED \cite{Myers:2003fd} by Myers and Pospelov which was later on extended to larger sections of the standard model by Bolokhov and Pospelov~\cite{Bolokhov:2007yc} and dimension 6 operators by Mattingly \cite{Mattingly:2008pw}.  The general set of higher dimension operators for free photons and fermions have recently been cataloged by Kostelecky and Mewes ~\cite{Kostelecky:2009zp, Kostelecky:2011gq}.

Why did all this attention to Lorentz breaking frameworks and observations develop in the late nineties and in the first decade of the new century? The answer is twofold as it is related to important developments coming from experiments and observation as well as from theoretical investigations.
Observationally, there were a number of puzzling observations related to gravity that spawned a corresponding growth in the zoo of quantum gravity models/scenarios with a low energy phenomenology. For example, in cosmology these are the years of the striking realization that our universe is undergoing an accelerated expansion phase \cite{Riess:1998cb,Perlmutter:1998np} which apparently requires a new exotic cosmological fluid, called dark energy, which violates the strong energy condition (to be added to the already well known, and still mysterious, dark matter component).

Also in the same period high energy astrophysics provided some new observational puzzles directly related to Lorentz symmetry, first with the apparent absence of the Greisen-Zatsepin-Kuzmin (GZK) cut off \cite{Greisen:1966jv,1969cora...11...45Z} (a suppression of the high-energy tail of the UHECR spectrum due to UHECR interaction with CMB photons) as claimed by the Japanese experiment AGASA \cite{Takeda:1998ps}, and later on with the so called TeV-gamma rays crisis, i.e. the apparent detection of a reduced absorption of TeV gamma rays emitted by AGN \cite{Protheroe:2000hp}. Both these ``crises" later on subsided or at least alternative, more orthodox explanations for them were advanced. However, their existence undoubtedly boosted the research in the field at that time.

It is perhaps this past ``training" that made several collaborations within the quantum gravity phenomenology community strongly emphasize the apparent incompatibility of the recent CERN--LNGS based experiment OPERA~\cite{Opera:2011zb} measurement of superluminal propagation of muonic neutrinos with Lorentz violating EFT (see e.g.~\cite{AmelinoCamelia:2011dx,Cohen:2011hx,Maccione:2011fr,Carmona:2011zg}. There is now evidence that the Opera measurement might be flawed due to unaccounted experimental errors and furthermore has been refuted by a similar measurement of the ICARUS collaboration~\cite{Antonello:2012hg}.  Nonetheless, this claim propelled a new burst of activity in Lorentz breaking phenomenology which might still provide useful insights for future searches.

Parallel to these exciting developments on the experimental/observational side, theoretical investigations provided new motivations for Lorentz breaking searches and constraints. Indeed, specific hints of LV arose from various approaches to Quantum Gravity. Among the many examples are the above mentioned string theory tensor VEVs \cite{KS89} and space-time foam models~\cite{AmelinoCamelia:1996pj,AmelinoCamelia:1997gz,Ellis:1999jf,Ellis:2000sx,Ellis:2003sd}, then semiclassical spin-network calculations in Loop QG~\cite{Gambini:1998it}, non-commutative geometry~\cite{Carroll:2001ws, Lukierski:1993wx, AmelinoCamelia:1999pm}, some brane-world backgrounds~\cite{Burgess:2002tb}. 

More recently, it cannot be omitted the role associated with the development of Lorentz breaking theories of gravity from early studies ~\cite{Gasperini:1985aw,Gasperini:1986xb,Gasperini:1987nq,Gasperini:1987fq,Gasperini:1998eb} to more systematic approaches such as Einstein--Aether \cite{Mattingly:2001yd,Eling:2004dk,Jacobson:2008aj} and Ho\v rava--Lifshitz  \cite{Horava:2009uw} gravity. Finally, there was the vigorous development over the same time of the so called condensed matter analogues of ``emergent gravity''~\cite{Barcelo:2005fc}, which showed how approximate local Lorentz invariance can arise from a fundamentally Galilean theory.  

Many of these approaches yield a low energy description of Lorentz violation in terms of effective field theory, and so before we delve into the specific operators that people have considered we remark on some generic aspects of embedding phenomenologically acceptable Lorentz violation in a low energy EFT.

\section{Modified dispersion relations and their naturalness}
\label{sec:TF}

We will concentrate here on the free field modifications to EFT from LV and hence talk primarily about modifications to dispersion relations.  This restriction isn't as limiting as it might seem.  For example, the necessary LV interaction terms generated when one wants to maintain gauge covariance in a LV theory~\cite{Colladay:1998fq} are controlled by the same LV coefficients that control the free theory so no new LV coefficients are introduced.  Since most experimental work is sensitive to free field behavior, constraints on those coefficients are generated from the phenomenology of the free part of the Lagrangian rather than the interaction part.  One can of course add other LV interaction terms by hand - such terms have not been as extensively studied in the literature and so we will not focus on them here. There is one caveat: when one sees constraints in the literature from modified particle decay mechanisms there is usually an assumption that the interactions of the standard model hold with only small modifications if any.  This is reasonable, as if there is only a tiny modification to the free field equations then by the argument above about how gauge generated LV interaction terms are controlled by the same coefficients as the free field part, the corresponding modifications to the interaction terms would also be small.  This would, in general, modify the rate of various reactions but the rate difference given the constraints detailed below would be unobservable.  
  
Turning back to the issues of modified dispersion relations in LV EFT, one has to firstly recognize that calculations of a specific dispersion modification from a specific quantum gravity theory are in general problematic and that cases where one can do so~\cite{AmelinoCamelia:1996pj} are the exception rather than the norm.   A more reasonable approach is to therefore simply consider a generic momentum expansion of a dispersion relation in a specified observer's frame of 
\begin{equation}
E^2=\vec{p}^2 + m^2 + \sum_{N=1}^\infty \eta^{(N)}_{\alpha_1\alpha2...\alpha_N} p^{\alpha_1} p^{\alpha_2}...p^{\alpha_N}
\end{equation}
where the low energy speed of light $c=1$ and each $\eta^{(N)}$ is an arbitrary rank N tensor with mass dimension $2-N$ and the $\alpha_i$ indices run over spacetime coordinates.  

Let us note that this ansatz assumes that the propagating mass eigenstates are also eigenstates of the Lorentz violating physics.  This does not need to be the case, and having the eigenstates of Lorentz violation not match the mass eigenstates can be useful when trying to analyze the effects of Lorentz violation on neutrinos.  However, since such a mismatch in other sectors would introduce oscillations and other unseen effects for particles other than neutrinos, i.e. those particles we are primarily concerned about.  Therefore we shall not consider such possibilities outside of the neutrino sector.  

The $\eta^{(N)}$'s can be mapped on to coefficients in a corresponding Lagrangian, although we note for the reader that in a generic Lagrangian there are many other coefficients that don't influence free field dispersion (c.f. ~\cite{Kostelecky:2009zp}).  Needless to say, as $N$ increases, there are a multitude of possible dispersion relations. Testing all such possible combinations can be done, of course, and would be the most systematic way to evaluate the possibility of Lorentz violation.  

One can, however, simplify the possible set of dispersion by making various assumptions such as CPT invariance, rotational invariance, etc. Rotational invariance is one of the most common assumptions made, for three reasons.~\footnote{A notable exception to this assumption is the Standard Model Extension and associated tests.  Rotational invariance is not assumed in the program of the standard model extension as it considers all terms at each mass dimension.}  The first is that it dramatically reduces the number of possible dispersion terms while still preserving interesting phenomenology.  Second, in many instances rotational invariance is more stringently tested than boost invariance so it provides a good model for testing boost invariance alone.  Finally, many QG models out there that do not predict exact Lorentz invariance still preserve rotational invariance. (See \cite{Jacobson:2005bg} for further discussion about this assumption.)  

With rotational invariance in mind in some frame, a common assumption for the dispersion relation is then
\begin{equation}%
E^2=p^2+m^2+\sum_{N=1}^{\infty} \tilde{\eta}_N p^N\;,%
\label{eq:disprel}%
\end{equation}%
where $p$ is the magnitude of the three momentum.  This type of expansion assumes high energy particles and that the corrections are small, so anywhere $E$ would have appeared on the RHS of \eqref{eq:disprel} it is replaced by $p$.  In general one can allow the LV parameters $\eta_N$ to depend on the particle type, and indeed it turns out that they {\it   must} sometimes be different but related in certain ways for photon polarization states, and for particle and antiparticle states, if the framework of effective field theory is adopted.   The lowest order LV terms ($p$, $p^2$, $p^3$, $p^4$) have been the terms that have generated the most attention (c.f. \cite{Mattingly:2005re, Kostelecky:2008ts} and references therein)\footnote{We disregard  here the possible appearance of dissipative terms \cite{Parentani:2007uq} in the dispersion relation, as this would correspond to a theory with unitarity loss and to a more radical departure from standard physics than that envisaged in the framework discussed herein (albeit a priori such dissipative scenarios are logically consistent and even plausible within some quantum/emergent gravity frameworks).}.

\subsection{The naturalness problem}

From a EFT point of view the only relevant operators should be the lowest order ones, i.e. those of mass dimension 3,4 corresponding to terms of order $p$ and $p^2$ in the dispersion relation. Situations in which higher order operators contribute to the dispersion as much as the lowest order ones at some energy are only possible at the cost of a severe, indeed arbitrary, fine tuning of the coefficients $\tilde{\eta}_N$ (which we discuss below). However, as we shall see current observational constraints are incredibly tight on dimension 3 operators and very severe on dimension 4 ones. This is kind of obvious, given that these operators would end up modifying the dispersion relation of elementary particles at easily achievable energies. Dimension 3 operators would dominate as $p\to 0$ while the dimension 4 ones would generically induce a species dependent, constant shift in the limit speed for elementary particles.  Hence one is left with two less than perfect approaches. 

First, one can assume the standard EFT hierarchy, stop testing at operators of mass dimension three and four, and, due to the tightness of limits, argue that Lorentz invariance is likely an exact symmetry of nature and that QG/emergent models that do not respect the symmetry should be discounted.  Or, one can avoid assumptions about whether any additional new physics comes into play between everyday energies and the QG scale and so not assume a particular hierarchy between operators of various mass dimension.  This is the approach many phenomenologists take: simply start at the lowest dimension operators, derive constraints, and work upwards in $N$ as far as one can observationally go without imposing any necessary hierarchy. 

Not assuming a necessary hierarchy \eqref{disprel2} and simply constraining the coefficients $\eta_N$ at each order is perfectly good phenomenologically, and we will take that approach going forward, but it is important that the reader understand why it is theoretically unnatural.   The reason such a hierarchy is unnatural is simple: in EFT radiative corrections will generically allow the percolation of higher dimension Lorentz violating terms into the lower dimension terms due to the interactions of particles~\cite{Collins:2004bp, Polchinski:2011za}. 

In EFT loop integrals will be naturally cut-off at the EFT breaking scale, if such scale is as well the Lorentz breaking scale  the two will effectively cancel leading to unsuppressed, coupling dependent, contributions to the base dimension four kinetic terms that generate the usual propagators.  Hence radiative corrections will not allow a dispersion relation with only $p^3$ or $p^4$ Lorentz breaking terms but will automatically induce extra unsuppressed LV terms in $p$ and $p^2$ which will be naturally dominant.  One could argue that RG effects might naturally suppress the sizes of these coefficients at low energies.  As we shall see, in specific models where the RG flow has been calculated, the running of LV coefficients is only logarithmic and so there is no indication that RG flow will actually drive coefficients to zero quickly in the infrared.

Several ideas have been advanced in order to resolve such a ``naturalness problem" (\see e.g.~\cite{Jacobson:2005bg}). While it would be cumbersome to review all the proposals here, we point out two of the more prominent ideas, both of which involve introducing a new scale into the problem in addition to the Lorentz violating scale.  If there are two scales $M$ and $\mu$ involved, there can be a hierarchy of LV coefficients different than the naive one, for example
\beq \tilde{\eta}_1=\eta_1 \frac{\mu^2}{M},\qquad
\tilde{\eta}_2=\eta_2 \frac{\mu}{M},\qquad \tilde{\eta}_3=\eta_3
\frac{1}{M}
\label{disprel2} \eeq
or
\beq \tilde{\eta}_2=\eta_2 \frac{\mu^2}{M^2},\qquad
\tilde{\eta}_4=\eta_4 \frac{1}{M^2}
\label{disprelnoodd} \eeq
where $M$ is the QG scale, usually taken to be the Planck scale, $\mu$ is some other far lower energy scale, and $\eta_N$ is a de-demensionalized coefficient usually assumed to be O(1).  The exact hierarchy, whether it involves terms of every mass dimension as in \eqref{disprel2}, or only even dimension as in \eqref{disprelnoodd} (which can be accomplished up to $N=4$ by imposing CPT) is model dependent.  We now briefly describe two ideas that have been put forward that would generate such scales.

\subsubsection{A new symmetry}

Most of the aforementioned proposals implicitly assume that the Lorentz breaking scale is the Planck scale. One then needs the EFT scale (which can be naively identified with what we called previously $\mu$) to be different from the Planck scale and actually sufficiently small so that the lowest order ``induced" coefficients can be suppressed by suitable small rations of the kind $\mu^p/M^q$ where $p,q$ are some positive powers.

A possible solution in this direction can be provided by introducing what is commonly called a ``custodial symmetry" - a symmetry other than Lorentz that forbids lower dimension operators and is broken at the low scale $\mu$. The most plausible candidate for this role is supersymmetry (SUSY)~\cite{GrootNibbelink:2004za,Bolokhov:2005cj}. SUSY is by definition a symmetry relating fermions to bosons  i.e.~matter with interaction carriers. As a matter of fact, SUSY is intimately related to Lorentz invariance. Indeed, it can be shown that the composition of at least two SUSY transformations induces space-time translations. However, SUSY can still be an exact symmetry even in presence of LV and can actually serve as a custodial symmetry preventing certain operators to appear in LV field theories. 

The effect of SUSY on LV is to prevent dimension $\leq 4$, renormalizable LV operators to be present in the Lagrangian.
Moreover, it has been demonstrated \cite{GrootNibbelink:2004za,Bolokhov:2005cj} that the renormalization group equations for Supersymmetric QED plus the addition of dimension 5 LV operators \`a la Myers \& Pospelov \cite{Myers:2003fd} do not generate lower dimensional operators if SUSY is unbroken. However, this is not the case for our low energy world, of which SUSY is definitely not a symmetry. 

The effect of soft SUSY breaking was also investigated in \cite{GrootNibbelink:2004za,Bolokhov:2005cj}. It was found there that, as expected, when SUSY is broken the renormalizable operators are generated. In particular, dimension $\kappa$ ones arise from the percolation of dimension $\kappa+2$ LV operators\footnote{We consider here only $\kappa = 3,4$, for which these relationships have been demonstrated.}. The effect of SUSY soft-breaking is, however, to introduce a suppression of order $m_{s}^{2}/M_{\rm Pl}$ ($\kappa=3$) or $(m_{s}/M_{\rm Pl})^{2}$ ($\kappa=4$), where $m_{s}\simeq 1$~TeV is the scale of SUSY soft breaking. Although, given present constraints, the theory with $\kappa=3$ needs a lot of fine tuning to be viable, since the SUSY-breaking-induced suppression is not enough powerful to kill linear modifications in the dispersion relation of electrons, if $\kappa = 4$ then the induced dimension 4 terms are suppressed enough, provided $m_{s} < 100$~TeV. Current lower bounds from the Large Hadron Collider are at most around 950 GeV for the most simple models of SUSY~\cite{ATLAS} (the so called ``constrained minimal supersymmetric standard model", CMSSM).

Finally, it is also interesting to note that the analogue model of gravity can be used as a particular implementation of the above mentioned mechanism for avoiding the so called naturalness problem via a custodial symmetry. This was indeed the case of multi-BEC~\cite{Liberati:2005pr,Liberati:2005id}. 

\subsubsection{Gravitational confinement of Lorentz violation}
\label{gravconf}

The alternative to an extra symmetry is to turn the problem upside down and posit that the Lorentz breaking scale (the $M$ appearing in the above dispersion relations) is not set by the Planck scale, but is instead the lower scale $\mu$.  If one does this and begins with a theory which has higher order Lorentz violating operators only in the gravitational sector, then one can hope that the gravitational coupling $G_N\sim M^{-2}_{\rm Pl}$ will suppress the ``percolation" to the matter sector where the constraints are strongest.  Matter Lorentz violating terms will all possess factors of the order $(\mu/M_{\rm Pl})^{2}$ which can become strong suppression factors if $\mu\ll M_{\rm Pl}$. This is the idea underlying the work presented in \cite{Pospelov:2010mp} which applies it to the special case of Ho\v rava--Lifshitz  gravity. There it was shown that indeed a workable low energy limit of the theory can be derived through this mechanism which apparently is fully compatible with existing constraints on Lorentz breaking operators in the matter sector.  In our opinion, this new route deserves further attention and should be more deeply explored in the future. 

\section{Dynamical Frameworks I:  Rotationally Invariant CPT even Standard Model Extension with mass dimension five and six operators}

We now turn from theoretical considerations about naturalness and retreat to the phenomenological approach of ``constrain everything as best one can observationally''. There are various systematic frameworks and approaches for this. The ``Standard Model Extension'' or SME contains all possible Lorentz violating tensors that can be coupled to standard model fields without changing the field content or violating gauge symmetry.  The standard model extension can be split into the ``Minimal Standard Model Extension''~\cite{KS89} of Kostelecky et. al.\/, which contains only renormalizable operators, and the full SME, which contains the infinite tower of non-renormalizable higher mass dimension operators.  As one can imagine, there are dozens of possible operators even for the renormalizable case, and the number of operators up to mass dimension six is in the hundreds.  Many of the operators can be constrained by similar methods, so for the purposes of this introduction we will concentrate on the most well studied sector of the SME, that of rotationally invariant QED.  In particular we will concentrate on the interactions of photons, electrons, and protons.

Usually when defining a field theory one starts with the renormalizable operators and proceeds in increasing mass dimension.  Here we start with higher mass dimension and work backwards for purely pedagogical reasons - many of the reactions and constraints we describe in detail for the higher dimension operators will also be useful for setting constraints on various lower dimension operators.  The first mass dimension we explore is mass dimension six, as this is the highest mass dimension where all the operators have been classified and significantly studied.  We break the operators into CPT even and odd classes as a different set of observations can be used to constraint CPT odd operators.

\subsection{The model}

A list of CPT even, rotationally invariant mass dimension five and six LV terms was computed in~\cite{Mattingly:2008pw} through the same procedure used by Myers \& Pospelov for dimension 5 LV (see below), and this has been extended to non-rotationally invariant operators in ~\cite{Kostelecky:2009zp, Kostelecky:2011gq}.  With rotation invariance all LV tensors must reduce to suitable products of a time-like vector field, usually denoted $u^{\alpha}$.  This is usually taken to be unit, so that in the frame of the observer whose world line is tangent to $u^\alpha$, $u^\alpha$ has components $(1,0,0,0)$.   This allows us to express constraints solely in terms of the numerical coefficient involved in any $u^\alpha$-matter interaction term.  Of course, the actual direction of $u^\alpha$ is technically arbitrary.  However, the common choice, which we make here, is to define $u^\a$ to be aligned with the rest frame of the cosmic microwave background.  In terms of $u^\a$ the known mass dimension six fermion operators are
\begin{eqnarray}
 - \frac{i}{\Mpl^{2}}\overline{\psi}(u\cdot D)^{3}(u\cdot \gamma)(\alpha_{L}^{(6)}P_{L} + \alpha_{R}^{(6)}P_{R}) \psi \\
 \nonumber-\frac{i}{\Mpl^{2}}\overline{\psi} (u\cdot D) \square (u\cdot \gamma) (\tilde{\alpha}_{L}^{(6)}P_{L} + \tilde{\alpha}_{R}^{(6)}P_{R}) \psi\;,
\label{eq:op-dim6-ferm}
\end{eqnarray}
where $P_{R,L}$ are the usual left and right spin projectors $P_{R,L} = (1\pm\gamma^{5})/2$ and $D$ is the gauge covariant derivative.  All the coefficients $\alpha$ are dimensionless because we factorize suitable powers of the Planck mass out explicitly.  In addition there is a CPT even dimension five term~\cite{Mattingly:2008pw},
\begin{equation}
-\frac{1} {\Mpl} \overline{\psi} (u \cdot D)^2 (\alpha^{(5)}_L P_L + \alpha^{(5)}_R P_R) \psi.
\end{equation}
The known photon operator is
\begin{equation}
-\frac{1}{2\Mpl^{2}}\beta_{\gamma}^{(6)}F^{\mu\nu}u_{\mu}u^{\sigma}(u\cdot\partial)F_{\sigma\nu}\;.
\label{eq:op-dim6-phot}
\end{equation}

From these operators, the dispersion relations of fermions and photons can be computed, yielding
\begin{eqnarray}
\nonumber 
E^{2} - p^{2} - m^{2} &=&   \frac{\alpha_{R}^{(6)} E^{3}}{\Mpl^{2}}(E+sp) + \frac{\alpha_{L}^{(6)}E^{3}}{\Mpl^{2}}(E-sp)+ \frac{m}{\Mpl}(\alpha_{R}^{(5)}+\alpha_{L}^{(5)})p^{2} + \alpha_{R}^{(5)}\alpha_{L}^{(5)}\frac{p^{4}}{\Mpl^{2}} \\
\omega^{2}-k^{2} &=& \beta^{(6)}\frac{k^{4}}{\Mpl^{2}}\;,
\label{eq:disp-rel-dimsix}
\end{eqnarray}
where $m$ is the electron mass and where $s = {\sigma}\cdot\mathbf{p}/|\mathbf{p}|$ is the helicity of the fermions. The $\tilde{\alpha}$ terms contribute as $m^2 /\Mpl^2$, i.e. highly suppressed, and so will be neglected. %However, this term is suppressed by the tiny ratio $m/\Mpl \sim 10^{-22}$ and can be safely neglected, provided that $E > \sqrt{m\Mpl}$.

In general, a LV dispersion for a particle with a certain set of quantum numbers (mass, spin, etc.) will be of the form $E^2=p^2+f^{(n)}p^n/\Mpl^{n-2}$, and so we will often refer to type ``n'' LV.  For example, because the high-energy fermion states are almost exactly chiral, we can further simplify the fermion dispersion relation eq.~(\ref{eq:disp-rel-dimsix}) (with $R=+$, $L=-$)
\begin{equation}
E^{2} = p^{2} + m^{2}  + \frac{m} {\Mpl} \eta^{(2)} p^{2} + \eta_{\pm}^{(4)} \frac{p^{4}}{\Mpl^{2}}\;
\label{eq:disp-rel-ferm-dim6-improved}
\end{equation}
where $\eta^{(n)}$ is the dispersion coefficient of the LV $p^n$ term in the dispersion relation for the fermion.  We choose $\eta$ as the coefficient as this nomenclature is common in the literature.  Similarly, $\xi^{(n)}$ will refer to the generic dispersion coefficient for a photon (so in the case above $\xi^{(4)}=\beta^{(6)}$).  As it is suppressed by a factor of order $m/\Mpl$, we will drop the quadratic modification generated by the dimension five operator. Indeed this can be safely neglected, provided that $E > \sqrt{m\Mpl}$. Let us stress however, that this is exactly an example of a dimension 4 LV term with a natural suppression, which for electron is of order $m_{e}/\Mpl \sim 10^{-22}$. Therefore, any limit larger than $10^{-22}$ placed on this term would not have to be considered as an effective constraint (to date, the best constraint for a rotationally invariant electron LV term of dimension 4 is $O(10^{-16})$ \cite{Stecker:2001vb}).  Note that modulo this the CPT even dimension five operator for fermions has the same effect on the dispersion as the CPT even dimension six in that it generates a $p^4$ term, so we will generally just write constraints directly on $\eta_{\pm}^{(6)}$.     It may seem puzzling that in a CPT invariant theory we distinguish between different fermion helicities in  \eqref{eq:disp-rel-ferm-dim6-improved}. However, although they are CPT invariant, some of the LV terms displayed in eq.~(\ref{eq:op-dim6-phot}) are odd under P and T. 

CPT invariance allows us to determine a relationship between the LV coefficients of fermions and anti-fermions. Indeed, to obtain these we simply realize that, by CPT, the dispersion relation of the anti-fermion is given by (\ref{eq:disp-rel-dimsix}), with the replacements $s \rightarrow -s$ and $p\rightarrow -p$. If $q,\overline{q}$ denote a charge fermion and anti-fermion, then the relevant anti-fermion coefficient $\eta^{(6)}_{\overline{q}}$ is such that $\eta^{(6)}_{\overline{q}_{\pm}} = \eta^{(6)}_{q_{\mp}}$, where $\overline{q}_{\pm}$ indicates an anti-fermion of positive/negative helicity (and similarly for the $q_{\pm}$). Let us anticipate that the same argument used above leads to the conclusions that for dispersion relations with odd powers of "n" (e.g.\/ $p^3$ type dispersion relations) one obtains  $\eta^{(n_{odd})}_{\overline{q}_{\pm}} = -\eta^{(n_{odd})}_{q_{\mp}}$. Hence, for arbitrary "n" one would expect  $\eta^{(n)}_{\overline{q}_{\pm}} = (-1)^n \eta^{(n)}_{q_{\mp}}$. This different behavior between even and odd powers "n" type dispersion relations leads to quite distinct phenomenologies as we shall see later.

\subsection{Constraints}

\subsubsection{Threshold reactions}
\label{sec:thresholds}

Threshold reactions with ultra-high energy cosmic rays provide the only significant constraints on the above operators.  A threshold reaction is a reaction that does not occur above a certain energy scale, which we call the ``threshold energy''.  An interesting and useful phenomenology of threshold reactions is introduced by LV in EFT; also, threshold theorems can be rederived \cite{Mattingly:2002ba}. Sticking to the present case of rotational invariance and monotonic dispersion relations (see \cite{Baccetti:2011us} for a generalization to more complex situations), the kinematics of threshold reactions yield a number of useful phenomenological facts about these reactions\cite{Jacobson:2002hd}:
\begin{itemize}
\item Threshold configurations still correspond to head-on incoming particles and parallel outgoing ones
\item The threshold energy of existing threshold reactions can shift, and upper thresholds (i.e.~maximal incoming momenta at which the reaction can happen in any configuration) can appear
\item Pair production can occur with unequal outgoing momenta
\item New, normally forbidden reactions can be viable
\end{itemize}

LV corrections are surprisingly important in threshold reactions because the LV term (which as a first approximation can be considered as an additional energy dependent ``mass'') should be compared not to the momentum of the involved particles, but rather to the (invariant) mass of the heaviest particle in the interaction. Thus, an estimate for the threshold energy is
\begin{equation}
p_{\rm th} \simeq \left(\frac{m^{2}\Mpl^{n-2}}{\eta^{(n)}}\right)^{1/n}\;,
\label{eq:threshold-general}
\end{equation}
where $m$ is the mass of the heaviest particle involved in the reaction. Interesting values for $p_{\rm th}$ are discussed, e.g., in \cite{Jacobson:2002hd} and given in Tab.~\ref{tab:thresholds}. %, from which it can be seen that r
\begin{table}[htbp]
\caption{Values of $p_{\rm th}$, according to eq.~(\ref{eq:threshold-general}), for different particles involved in the reaction: neutrinos, electrons and proton. Here we assume $\eta^{(n)} \simeq 1$.}
\begin{center}
\begin{tabular}{|c|c|c|c|}
\hline
& $m_{\nu}\simeq 0.1~\eV$ & $m_{e}\simeq 0.5~\MeV$ & $m_{p} \simeq 1~\GeV$ \\
\hline
$n=2$ & 0.1 eV & 0.5 MeV & 1 GeV\\
\hline 
$n=3$ & 500 MeV & 14 TeV & 2 PeV\\
\hline 
$n=4$ & 33 TeV & 74 PeV & 3 EeV\\
\hline
\end{tabular}
\end{center}
\label{tab:thresholds}
\end{table}%
Reactions involving neutrinos are the best candidate for observation of LV effects, whereas electrons and positrons can provide results for $n=3$ theories but cannot readily be accelerated by astrophysical objects up to the required energy for $n=4$. In this case reactions of protons can be very effective, because cosmic rays can have measured energies well above 3 EeV. We now discuss two threshold reactions are of particular use when constraining $n=4$ LV.  \\
\\
\textit{LV-allowed threshold reactions: $\gamma$-decay} \\
\\
The decay of a photon into an electron/positron pair is made possible by LV because energy-momentum conservation may now allow otherwise forbidden reactions to occur. Since the decay is a reaction described by the fundamental QED vertex, the rate once above threshold will be quite fast.   The threshold for this process, is set by the condition \cite{Jacobson:2005bg}
\begin{equation}
k_{th} \approx \left( \frac{m^2\Mpl^{n-2}} {(F(\eta^{(n)},\xi^{(n)}))^{n-2}} \right)^{1/n}\, ,
\end{equation}
where $F(\eta^{(n)},\xi^{(n)})$ is a linear combination of $\eta^{(n)},\xi^{(n)}$.  Notably, the electron-positron pair can be created with slightly different outgoing momenta (asymmetric pair production).  Furthermore, the decay rate is extremely fast above threshold \cite{Jacobson:2005bg} and is of the order of $(10~{\rm ns})^{-1}$ ($n=3$) or $(10^{-6}~{\rm ns})^{-1}$ ($n=4$) if the LV coefficients are of O(1).\\
%\item[Vacuum \v{C}erenkov]
\\
\textit{LV-allowed threshold reactions: Vacuum \v{C}erenkov} \\
\\
In the presence of LV, the process of Vacuum \v{C}erenkov (VC) radiation $q^{\pm}\rightarrow q^{\pm}\gamma$, where $q$ is a charged fermion can occur as this is just a rotated diagram of $\gamma$-decay. The threshold energy of the reaction is roughly the same and so is also given by

\begin{equation}
\label{eq:energythreshvc}
E_{th} \approx \left( \frac{m^2\Mpl^{n-2}} {(F(\eta^{(n)},\xi^{(n)})^{n-2}} \right)^{1/n}\, . 
\end{equation}

Just above threshold this process is also an extremely efficient method of energy loss.  Note that while $\gamma$ decay destroys the incoming photon, the vacuum \v{C}erenkov effect merely is an energy loss process.\\
\\
%\subsubsection{Photon absorption}
\textit{LV-modified threshold reactions: Photon Absorption}\\
\\
A process related to photon decay is photon absorption, $\g\g\rightarrow e^+e^-$. Unlike photon decay,
this is allowed in Lorentz invariant QED and it plays a crucial role in making our universe opaque to gamma rays above tens of TeVs. 

If one of the photons has energy $\omega_0$, the threshold for the reaction occurs in a
head-on collision with the second photon having the momentum
(equivalently energy) $k_{\rm LI}=m^2/\omega_{0}$. For example, if $k_{\rm LI}=10$ TeV (the typical energy of inverse Compton generated photons in some active galactic nuclei) the
soft photon threshold $\omega_0$ is approximately 25 meV, corresponding to a wavelength of 50 microns.

In the presence of Lorentz violating dispersion relations the threshold for this process is in general altered, and the process
can even be forbidden. Moreover, as firstly noticed by Klu\'zniak~\cite{Kluzniak:1999qq}, in some cases there is an upper
threshold beyond which the process does not occur. Physically, this means that at sufficiently high momentum the photon does not carry enough energy to create a pair and simultaneously conserve energy and momentum. Note also, that an upper threshold can only be found in regions of the parameter space in which the $\gamma$-decay is forbidden, because if a single photon is able to create a pair, then {\em a fortiori} two interacting photons will do \cite{Jacobson:2002hd}. 

Let us exploit the above mentioned relation $\eta_{\pm}^{e^{-}} = (-)^{n}\eta_{\mp}^{e^{+}}$ between the electron-positron coefficients, and assume that on average the initial state is unpolarized. In this case, using the energy-momentum conservation, the kinematics equation governing pair production is the following \cite{Jacobson:2005bg}
\begin{equation}
\frac{m^{2}}{k^{n}y \left(1-y\right)} =  
  \frac{4\omega_{b}}{k^{n-1}} + \tilde{\xi} - \tilde{\eta} \left( y^{n-1}+(-)^{n}\left(1-y\right)^{n-1}\right)
   \label{eq:ggscat}
\end{equation}
where $\tilde{\xi}\equiv\xi^{(n)}/M^{n-2}$ and $\tilde{\eta}\equiv\eta^{(n)}/M^{n-2}$ are respectively the photon's and electron's LV coefficients divided by powers of $M$, $0 < y < 1$ is the fraction of momentum carried by either the electron or the positron with respect to the momentum $k$ of the incoming high-energy photon and $\omega_{b}$ is the energy of the target photon. The  analysis is more complicated than simple one-particle initial state decay or radiative processes. In particular it becomes necessary to sort out whether the thresholds are lower or upper ones and whether they occur with the equal or different pair momenta.  
%However, one can perform a complete threshold analysis in this case and derive useful constraints~\cite{Jacobson:2002hd}.

\subsubsection{Constraints from GZK secondaries}

One of the most interesting features related to the physics of Ultra-High-Energy Cosmic Rays (UHECRs) is the Greisen-Zatsepin-Kuzmin (GZK) cut off \cite{Greisen:1966jv,1969cora...11...45Z}, a suppression of the high-energy tail of the UHECR spectrum arising from interactions with CMB photons, according to $p\gamma\rightarrow \Delta^{+}\rightarrow p\pi^{0}(n\pi^{+})$. This process has a (LI) threshold energy $E_{\rm th} \simeq 5\times 10^{19}~(\omega_{b}/1.3~\meV)^{-1}~\eV$ ($\omega_{b}$ is the target photon energy). Experimentally, the presence of a suppression of the UHECR flux was claimed only recently \cite{Abbasi:2007sv,Roth:2007in}. Although the cut off could be also due to the finite acceleration power of the UHECR sources, the fact that it occurs at the expected energy favors the GZK explanation. The results presented in \cite{Cronin:2007zz} seemed to further strengthen this hypothesis (but see further discussion below).

Rather surprisingly, significant limits on $\xi=\xi^{(6)}$ and $\eta=\eta^{(6)}$ for the proton can be derived by considering UHE photons generated as secondary products of the GZK reaction\cite{Galaverni:2007tq,Maccione:2008iw}. This can be used to further improve the constraints on dimension 5 LV operators and provide a first robust constraint of QED with dimension 6 CPT even LV operators. 

These UHE photons originate because the GZK process leads to the production of neutral pions that subsequently decay into photon pairs. These photons are mainly absorbed by pair production onto the CMB and radio background. Thus, the fraction of UHE photons in UHECRs is theoretically predicted to be less than 1\% at $10^{19}~\eV$ \cite{Gelmini:2005wu}. Several experiments imposed limits on the presence of photons in the UHECR spectrum. In particular, the photon fraction is less than 2.0\%, 5.1\%, 31\% and 36\% (95\% C.L)~at $E = 10$, 20, 40, 100 EeV  respectively \cite{Aglietta:2007yx,Rubtsov:2006tt}. 

However, we have just seen that pair production can be strongly affected by LV. In particular, the (lower) threshold energy can be slightly shifted and in general an upper threshold can be introduced \cite{Jacobson:2002hd}. If the upper threshold energy is lower than $10^{20}~\eV$, then UHE photons are no longer attenuated by the CMB and can reach the Earth, constituting a significant fraction of the total UHECR flux and thereby violating experimental limits \cite{Galaverni:2007tq,Maccione:2008iw,Galaverni:2008yj}. 

Moreover, it has been shown \cite{Maccione:2008iw} that the $\gamma$-decay process can also imply a significant constraint. Indeed, if some UHE photon ($E_{\gamma}\simeq 10^{19}~\eV$) is detected by experiments (and the Pierre Auger Observatory, PAO, will be able to do so in few years \cite{Aglietta:2007yx}), then $\gamma$-decay must be forbidden above $10^{19}~\eV$. 

\begin{figure}[tbp]
\sidecaption[t]
 \includegraphics[scale=0.4, angle = 90]{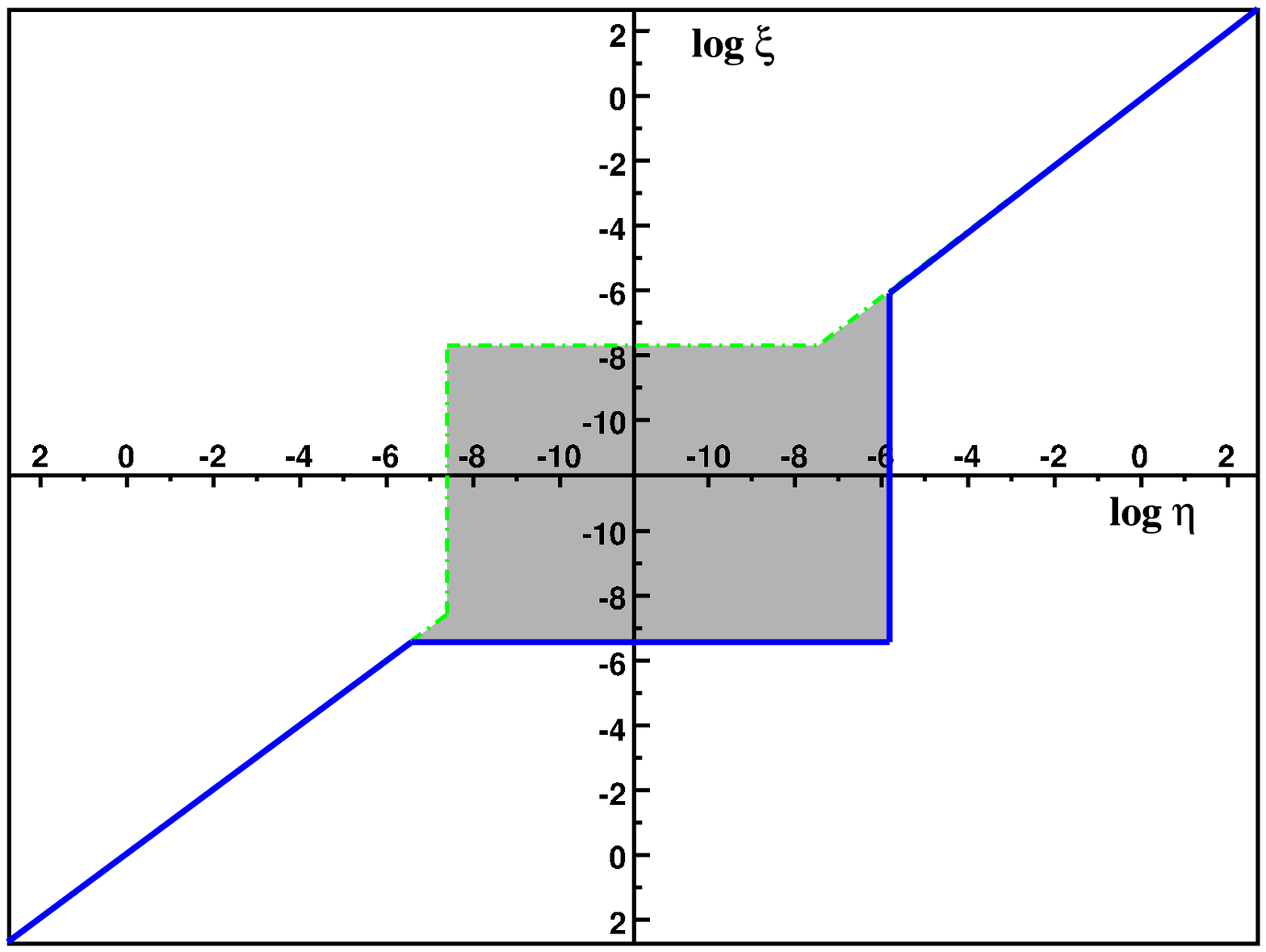}
 \caption{
 %Left panel: LV induced by dimension 5 operators. Right panel: 
 LV induced by dimension 6 operators. The LV parameter space is shown. The allowed regions are shaded grey. Green dotted lines represent values of $(\eta, \xi)$ for which the $\gamma$-decay threshold $k_{\gamma-dec} \simeq 10^{19}~\eV$. Solid, blue lines indicate pairs $(\eta,\xi)$ for which the pair production upper threshold $k_{\rm up} \simeq 10^{20}~\eV$. Picture taken from \cite{Liberati:2009pf}.}
 \label{fig:constraints}
\end{figure}

In conclusion we show in Fig.~\ref{fig:constraints} the overall picture of the constraints of QED dimension 6 LV operators, where the green dotted lines do not correspond to real constraints, but to the ones that will be achieved when AUGER will observe, as expected, some UHE photon.

\section{Dynamical Frameworks II:  Rotationally Invariant CPT odd Standard Model Extension with mass dimension five operators}

\subsection{The model}
Myers \& Pospelov \cite{Myers:2003fd} found that there are essentially only three operators of dimension five, quadratic in the fields, that can be added to the QED Lagrangian that give rise to dispersion modifications of type $n=3$.\footnote{Actually these criteria allow the addition of other (CPT even) terms, but these would not lead to modified dispersion relations (they can be thought of as extra, Planck suppressed, interaction terms) \cite{Bolokhov:2007yc}.} 
These extra-terms are the following:
\begin{equation}
-\frac{\xi}{2\Mpl}u^mF_{ma}(u\cdot D)(u_n\tilde{F}^{na}) + \frac{1}{2\Mpl}u^m\overline{\psi}\gamma_m(\zeta_1+\zeta_2\gamma_5)(u\cdot D)^2\psi\:,
\label{eq:LVterms}
\end{equation}
where $\tilde{F}$ is the dual of $F$ and $\xi$, $\zeta_{1,2}$ are dimensionless parameters. All these terms also violate CPT symmetry.
  More recently, this construction has been extended to the whole SM \cite{Bolokhov:2007yc, Kostelecky:2009zp,Kostelecky:2011gq}. 

From (\ref{eq:LVterms}) the dispersion relations of the fields are
modified as follows. For the photon one has
\begin{equation}
\omega_{\pm}^2 = k^2 \pm \frac{\xi^{(3)}}{\Mpl}k^3\:,
\label{eq:disp_rel_phot}
\end{equation}
where $\xi^{(3)}=\xi$ and the $+$ and $-$ signs denote right and left circular polarisation, while
for the fermion (with the $+$ and $-$ signs now denoting positive and
negative helicity states% 
)
\begin{equation}
E_\pm^2 = p^2 + m^2 + \eta_\pm^{(3)} \frac{p^3}{\Mpl}\  ;,
\label{eq:disp_rel_ferm}
\end{equation}
with $\eta_\pm^{(3)}=2(\zeta_1\pm \zeta_2)$. For the anti-fermion, it can be
shown by simple ``hole interpretation" arguments that the same
dispersion relation holds, with $\eta^{(3),af}_\pm = -\eta^{(3),f}_\mp$ where
$af$ and $f$ superscripts denote respectively anti-fermion and
fermion coefficients~\cite{Jacobson:2005bg,Jacobson:2003bn}.   Note that if CPT ends up being a fundamental symmetry of nature without Lorentz symmetry it would forbid all of the above mentioned CPT odd operators.  

\subsection{An aside on naturalness}
With the specific dimension five operators in hand, we can now return to a previously mentioned point, that RG flow does not significantly suppress the sizes of operators at low energies.  Let us consider the evolution of the dimension five LV parameters.  Bolokhov \& Pospelov \cite{Bolokhov:2007yc} addressed the problem of calculating the renormalization group equations for QED and the Standard Model extended with dimension-five operators that violate Lorentz Symmetry. In the framework defined above, assuming that no extra physics enters between the low energies at which we have modified dispersion relations and the Planck scale at which the full theory is defined, the evolution equations for the LV terms in eq.~(\ref{eq:LVterms}) that produce modifications in the dispersion relations, can be inferred as
\begin{equation}
%%\fl
\label{eq:RG}
\frac{d\zeta_1}{dt} =  \frac{25}{12}\,\frac{\alpha}{\pi}\,\zeta_1\; , \quad
\frac{d\zeta_2}{dt} =  \frac{25}{12}\,\frac{\alpha}{\pi}\,\zeta_2 - \frac{5}{12}\,\frac{\alpha}{\pi}\,\xi\; , \quad
\frac{d\xi}{dt} =  \frac{1}{12}\,\frac{\alpha}{\pi}\,\zeta_2 - \frac{2}{3}\,\frac{\alpha}{\pi}\,\xi \;,
\end{equation}
where $\alpha = e^2/4\pi \simeq 1/137$ ($\hbar = 1$) is the fine structure constant and $t = \ln(\mu^2/\mu_0^2)$ with $\mu$ and $\mu_0$ two given energy scales. (Note that the above formulae are given to lowest order in powers of the electric charge, which allows one to neglect the running of the fine structure constant.)

These equations show that the running is only logarithmic and therefore low energy constraints are robust: $O(1)$ parameters at the Planck scale are still $O(1)$ at lower energy. Moreover, they also show that $\eta^{(3)}_{+}$ and $\eta^{(3)}_{-}$ cannot, in general, be equal at all scales.  Similar calculations in the context of the renormalizable SME give equivalent results.

\subsection{Constraints}

We now detail some of the constraints that can be put on the CPT odd dimension five operators.  For a thorough review of these constraints, see also~\cite{Liberati:2009pf}.

\subsubsection{Photon time of flight}
\label{subsec:tof}

Although photon time-of-flight constraints from high energy photons propagating from cosmologically distant objects currently provide limits several orders of magnitude weaker than other constraints, they have been widely adopted in the astrophysical community. They were one of the first to be proposed in the seminal paper \cite{AmelinoCamelia:1997gz}.  More importantly, given their purely kinematical nature, they may be applied to a broad class of frameworks, even beyond EFT with LV. 

In general, a photon dispersion relation in the form of (\ref{eq:disp_rel_phot}) implies that photons of different colors (wave vectors $k_1$ and $k_2$) travel at slightly different speeds.  Let us first ignore any birefringence effects, and just consider some coefficient $\xi^{(n)}$ that is universal for all photons.  
Then, upon propagation on a cosmological distance $d$, the effect of energy dependence of the photon group velocity produces a time delay
\begin{equation}
 \Delta t^{(n)} = \frac{n-1}{2}\, \frac{k_2^{n-2}-k_1^{n-2}}{\Mpl^{n-2}}\,\xi^{(n)}\, d\;,
\label{eq:tof-naive}
\end{equation}
which clearly increases with $d$ and with the energy difference as long as $n>2$. The largest systematic error affecting this method is the uncertainty about whether photons of different energy are produced simultaneously in the source.% (e.g.,~a Gamma-Ray Burst, or an Active Galactic Nucleus).

One way to alleviate systematic uncertainties is to apply EFT, which gives more information than just a modified dispersion for photons.  In particular, one knows that the term generated by $\xi^{(3)}$,  in an EFT context implies birefringence.  Furthermore, photon beams generally are not circularly polarized; thus, they are a superposition of super and subluminal circularly polarized modes. Hence one can remove any systematic uncertainty relating to source dynamics or measured energy by measuring the velocity difference between the two polarization states at a single energy, corresponding to
\begin{equation}
 \Delta t = 2|\xi^{(3)}|k\, d/\Mpl\;.
\end{equation}
This bound would require that both polarizations be observed and that no spurious helicity-dependent mechanism (such as, for example, propagation through a birefringent medium) affects the relative propagation of the two polarization states.  

Note that one doesn't have to actually measure the actual polarization - a double peak in the high energy spectra of GRB's with a separation that scaled linearly with distance would be a smoking gun for birefringent theories.   However, if the polarization states do not fully separate then extracting a signal becomes more complicated.  Since current time of flight constraints compare low to high energy photons, birefringence destroys this.  A GRB burst would not have low energy photons arriving first with high energy following (or vice vera).  Instead the structure of the burst for a birefringent theory would be high energy photons first, low energy photons following, and then more high energy photons at the end.  Therefore, the net effect of this superposition may be to partially or completely erase the time-delay effect as it is usually calculated. 

In order to compute this modulating effect on a generic photon beam in a birefringent theory, let us describe a beam of light by means of the associated electric field, and let us assume that this beam has been generated with a Gaussian width
\begin{equation}
\vec{E} = A\, \left(e^{i(\Omega_{0}t-k^{+}(\Omega_{0})z)}\,e^{-(z-v_{g}^{+}t)^{2}\delta\Omega_{0}^{2}}\hat{e}_{+} + e^{i(\Omega_{0}t-k^{-}(\Omega_{0})z)}\,e^{-(z-v_{g}^{-}t)^{2}\delta\Omega_{0}^{2}}\hat{e}_{-}       \right)\;,
\end{equation}
where $\Omega_{0}$ is the wave frequency, $\delta\Omega_{0}$ is the gaussian width of the wave, $k^{\pm}(\Omega_{0})$ is the ``momentum'' corresponding to the given frequency according to (\ref{eq:disp_rel_phot}) and $\hat{e}_{\pm}\equiv (\hat{e}_{1}\pm i\hat{e}_{2})/\sqrt{2}$ are the helicity eigenstates. Note that by complex conjugation $\hat{e}_{+}^{*} = \hat{e}_{-}$. Also, note that $k^{\pm}(\omega) = \omega \mp \xi \omega^{2}/\Mpl$. Thus,
\begin{equation}
\vec{E} = A\, e^{i\Omega_{0}(t-z)}\left(e^{i\xi\Omega_{0}^{2}/\Mpl z}\,e^{-(z-v_{g}^{+}t)^{2}\delta\Omega_{0}^{2}}\hat{e}_{+} + e^{-i\xi\Omega_{0}^{2}/\Mpl z}\,e^{-(z-v_{g}^{-}t)^{2}\delta\Omega_{0}^{2}}\hat{e}_{-}       \right)\;.
\end{equation}
The intensity of the wave beam can be computed as
\begin{eqnarray}
\nonumber \vec{E}\cdot\vec{E}^{*} &=& |A|^{2}\left( e^{2i\xi\Omega_{0}^{2}/\Mpl z} + e^{-2i\xi\Omega_{0}^{2}/\Mpl z}  \right) e^{-\delta\Omega_{0}^{2}\left( (z-v_{g}^{+}t)^{2} + (z-v_{g}^{-}t)^{2}\right)}\\
&=& 2|A|^{2}e^{-2\delta\Omega_{0}^{2}(z-t)^{2}}\cos\left( 2\xi\frac{\Omega_{0}}{\Mpl}\Omega_{0}z\right)e^{ - 2\xi^{2}\frac{\Omega_{0}^{2}}{M^{2}}(\delta\Omega_{0}t)^{2}}\;.
\end{eqnarray}
This shows that there is an effect even on a linearly-polarised beam. The effect is a modulation of the wave intensity that depends quadratically on the energy and linearly on the distance of propagation. In addition, for a gaussian wave packet, there is a shift of the packet centre, that is controlled by the square of $\xi^{(3)}/\Mpl$ and hence is strongly suppressed with respect to the cosinusoidal modulation. Hence by looking for modulation of the signal with energy and distance one can in principle determine if the LV is birefringent from time of flight information even if the photons arrive as part of one ``burst''.

So far, the most robust constraints on $\xi^{(3)}$, derived from time of flight differences, have been obtained within the $D-$brane model (discussed in section \ref{sec:nonEFT}) from a statistical analysis applied to the arrival times of sharp features in the intensity at different energies from a large sample of GRBs with known redshifts~\cite{Ellis:2005wr}, leading to limits $\xi^{(3)}\leq O(10^3)$.
The importance of systematic uncertainties can be found in \cite{Albert:2007qk}, where the strongest limit $f^{(3)} < 47$ is found by looking at a very strong flare in the TeV band of the AGN Markarian 501.    Finally, an extremely strong limit (for this method at least) of $\xi^{(3)}<0.8$ has been obtained from the short, high energy gamma ray burst GRB 090510~\cite{Ackermann:2009aa}.  %Furthermore, by looking at the flare time-structure, the authors also claimed a best fit for $\xi^{(3)} \sim O(1)$ (note, however, that a similar fit could also be achieved by standard plasma physics).

\subsubsection{Photon polarization}
%\subsubsection{Polarization of low energy photons}
\label{sec:birefringence}

%While one cannot (yet) get reliable polarization measurements from GRB's, one can achieve good polarizations at much lower energies.  
Since electromagnetic waves with opposite circular polarizations have slightly different group velocities in rotationally invariant EFT LV when CPT is violated, the polarization vector of a linearly polarised plane wave with energy $k$ rotates. During the wave propagation over a distance $d$, the rotation angle for $n=3$ dispersion modifications is \cite{Jacobson:2005bg}
\footnote{Note that for an object located at cosmological distance (let $z$ be its redshift), the distance $d$ becomes
\begin{equation}
d(z) = \frac{1}{H_{0}}\int^{z}_0 \frac{1+z'}{\sqrt{\Omega_{\Lambda} + \Omega_{m}(1+z')^{3}}}\,dz'\;,
\end{equation}
where $d(z)$ is not exactly the distance of the object as it includes a $(1+z)^{2}$ factor in the integrand to take into account the redshift acting on the photon energies.}
\begin{equation} 
\theta(d) = \frac{\omega_{+}(k)-\omega_{-}(k)}{2}d \simeq \xi^{(3)}\frac{k^2 d}{2\,M_{\rm Pl}}\;.
\label{eq:theta}
\end{equation} 

Observations of polarized light from a distant source can then lead to a constraint on $|\xi^{(3)}|$ that, depending on the amount of available information --- both on the observational and on the theoretical (i.e.~astrophysical source modeling) side --- can be cast in two different ways \cite{Maccione:2008tq}:
\begin{enumerate}
\item
%\underline{Decrease in polarization degree}: 
Because detectors have a finite energy bandwidth, Eq.~(\ref{eq:theta}) is never probed in real situations. Rather, if some net amount of polarization is measured in the band $k_{1} < E < k_{2}$, an order-of-magnitude constraint arises from the fact that if the angle of polarization rotation (\ref{eq:theta}) differed by more than $\pi/2$ over this band,
the detected polarization would fluctuate sufficiently for the net signal polarization to be suppressed \cite{Gleiser:2001rm, Jacobson:2003bn}.  %%%% slightly different meaning
%As the difference in the rotation angle over the
%energy range is
From (\ref{eq:theta}), this constraint is
\begin{equation} 
\xi^{(3)}\lesssim\frac{\pi\,M_{\rm Pl}}{(k_2^2-k_1^2)d(z)}\;,
\label{eq:decrease_pol}
\end{equation} 
% 
%the constraint corresponds to impose $\Delta\theta \leq \pi/2$.  
This constraint requires that any intrinsic polarization (at source) not be
completely washed out during signal propagation. It thus relies on the
mere detection of a polarized signal; there is no need to consider the observed
polarization degree.
A more refined limit can be obtained by calculating the maximum
observable polarization degree, given the maximum intrinsic value \cite{McMaster}:
\begin{equation} 
\Pi(\xi) = \Pi(0) \sqrt{\langle\cos(2\theta)\rangle_{\mathcal{P}}^{2}
+\langle\sin(2\theta)\rangle_{\mathcal{P}}^{2}},
\label{eq:pol}
\end{equation} 
where $\Pi(0)$ is the maximum intrinsic degree of polarization,
$\theta$ is defined in eq.~(\ref{eq:theta}) and the average is
weighted over the source spectrum and instrumental efficiency,
represented by the normalized weight function
$\mathcal{P}(k)$~\cite{Gleiser:2001rm}.  
%%%%
Conservatively, one can set $\Pi(0)=100\%$, but a lower value 
may be justified on the basis of source modeling.
Using \eqref{eq:pol}, one can then 
cast a constraint by %%%% estimating how large $\xi$ can be so not to
%lower $\Pi$ below the observed value.
requiring $\Pi(\xi)$ to exceed the observed value. 

\item
%%%%\underline{Rotation of polarization angle}: 
%%%% Let us suppose 
Suppose 
%%%%
that
polarized light %%%% has been 
%%%%
measured in a certain energy band %%%% with 
has
%%%%
a position angle $\theta_{\rm obs}$ with respect to a fixed
direction. At fixed energy, the polarization vector rotates by the
angle (\ref{eq:theta}) \footnote{Faraday rotation is negligible at
these energies.}; if the position angle is measured by averaging over a
certain energy range, the final net rotation 
%%%% 
$\left<\Delta\theta\right>$
is given by the
superposition of the polarization vectors of all the photons in that
range:
%%%%
%
%. Thus, if $\theta_{\rm f}$ is the position angle after
%propagation, referred to the position angle at emission $\theta_{\rm
%i}$, the following relation holds
%
\begin{equation}
\tan (2\left\langle\Delta\theta\right\rangle) = \frac{
\left\langle\sin(2\theta)\right\rangle_{\mathcal{P}}}{\left\langle
\cos(2\theta)\right\rangle_{\mathcal{P}}}\;,
\label{eq:caseB}
\end{equation}
where %%%% $\theta = \xi(d/2M)k^{2}$ is the angle of rotation for a photon
%of energy $k$ with respect to $\theta_{\rm i}$.
$\theta$ is given by (\ref{eq:theta}).
If the position angle at emission $\theta_{\rm i}$ in the same energy band %%%% can be determined by
is known from a model of the emitting source, a constraint can be set by imposing
\begin{equation}
\tan(2\left\langle\Delta\theta\right\rangle) < \tan(2\theta_{\rm obs}-2\theta_{\rm i})\;.
\label{eq:constraint-caseB}
\end{equation}
%
%%%% rephrasing
Although this limit is tighter than those based on Eqs.~(\ref{eq:decrease_pol}) and (\ref{eq:pol}), it clearly hinges on assumptions about the 
%%%% process
%of emission of polarized light
nature of the source, 
%%%%
which may introduce significant uncertainties.
\end{enumerate}

\begin{figure}[tbp]
\sidecaption[t]
\includegraphics[scale=0.4]{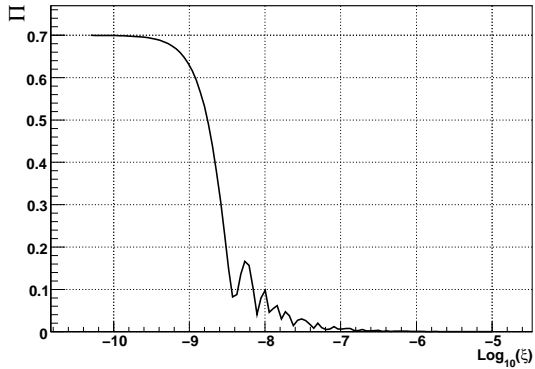} % requires the graphicx package
\caption{Constraint for the polarization degree. Dependence of $\Pi$
on $\xi$ for the distance of the CN and photons in the 150--300 keV
range, for a constant instrumental sensitivity $\mathcal{P}(k)$. Picture taken from \cite{Maccione:2008tq}.}
\label{fig:caseA}
\end{figure}

In the case of the Crab Nebula, a $(46\pm10)$\%
degree of linear polarization in the $100~\keV - 1~\MeV$ band has 
recently been measured 
%%%% estimated for the emission from the CN 
by the INTEGRAL mission
\cite{integral,integralpol}.  This 
%%%% rephrasing
measurement uses all photons within the SPI instrument energy band. However
the convolution of the instrumental sensitivity to polarization with
the detected number counts as a function of energy, $\mathcal{P}(k)$,
is maximized and approximately constant within a narrower energy band
(150 to 300 keV) and falls steeply outside this range \cite{McGlynn:2007pz}.  For this reason we shall,
conservatively, assume that most polarized photons are concentrated in
this band.
%%%% \underline{Decrease in polarization degree}: 
Given $d_{\rm Crab}=1.9~\kpc$, $k_2 = 300~\keV$ and $k_1 = 150~\keV$, 
%%%% imposing $\Delta\theta < \pi/2$ in 
eq.~(\ref{eq:decrease_pol}) leads to the
order-of-magnitude estimate $|\xi| \lesssim 2\times10^{-9}$.
%Remember: $1=2\times 10^{-7}$ eV m.
%%%% Rephrasing
A more accurate limit follows from 
(\ref{eq:pol}).
%relies on the fact that the measured
%polarization degree cannot be lower than $\sim46 \pm 10 \%$ (see 
%eq.~(\ref{eq:pol})). While the most conservative assumption would be to
%consider an intrinsic $100\%$ polarization level, 
In the case of the CN there is
a robust understanding that photons in the range of interest are
produced via the synchrotron process, for which the maximum degree of
intrinsic linear polarization is about $70\%$ (see e.g.~\cite{Petri:2005ys}).
Figure \ref{fig:caseA} illustrates the dependence of $\Pi$ on $\xi$ (see Eq.~\eqref{eq:pol})
for the distance of the CN and for $\Pi(0)=70\%$. 
The requirement $\Pi(\xi)>16\%$
(taking account of a $3\sigma$ offset from the best fit value $46\%$) leads to
the constraint (at 99\% CL)
%and shows that the requirement that an
%intrinsic polarization level $~70 \%$ is not lowered below 
%
\begin{equation}
|\xi| \lesssim 6\times 10^{-9}\;.
\label{eq:constraint-degree}
\end{equation}
%%%% Incidentally, 
It is interesting to notice that X-ray polarization
measurements of the CN already available in 1978
\cite{1978ApJ...220L.117W}, set a constraint $|\xi| \lesssim
5.4\times10^{-6}$, only one order of magnitude less stringent than
that reported in \cite{Fan:2007zb}.

%%%% \underline{Rotation of polarization angle}: 
%%%% rephrasing
Constraint (\ref{eq:constraint-degree}) can be tightened
by exploiting the current astrophysical understanding of the source.
The CN is a cloud of relativistic particles and fields powered
by a rapidly rotating, strongly magnetized neutron star. Both the 
Hubble Space Telescope and the Chandra X-ray satellite have
imaged the system, revealing a jet and torus that clearly identify the
neutron star rotation axis \cite{Ng:2003fm}. The projection of this
axis on the sky lies at a position angle of
$124.0^{\circ}\pm0.1^{\circ}$ (measured from North in
anti-clockwise). The neutron star itself emits pulsed radiation at its
rotation frequency of 30 Hz. In the optical band these pulses are
superimposed on a fainter steady component with a linear polarization
degree of ~30\% and direction precisely aligned with that of the
rotation axis \cite{Kanbach:2005kf}.  The direction of polarization
measured by INTEGRAL-SPI in the $\gamma$-rays is $\theta_{\rm obs} =
123^{\circ}\pm11^{\circ}$ ($1\sigma$ error) from the North, thus also
closely aligned with the jet direction and remarkably consistent with
the optical observations.

This compelling (theoretical and observational) evidence allows us to
use eq.~(\ref{eq:constraint-caseB}). Conservatively assuming
%%%% eliminated $\Delta\theta$ to avoid confusion
$\theta_{\rm i}-\theta_{\rm obs} = 33^{\circ}$
(i.e.~$3\sigma$ from $\theta_{\rm i}$, 99\% CL), this translates into
the limit
\begin{equation}
|\xi^{(3)}| \lesssim 9\times10^{-10}\;,
\label{eq:constraint-serious-crab}
\end{equation}
and $|\xi^{(3)}| \lesssim 6\times10^{-10}$ for a $2\sigma$ deviation (95\%
CL). 
%%%% no new paragraph

Polarized light from GRBs has also been detected and given their cosmological distribution they could be ideal sources for improving the above mentioned constraints from birefringence. Attempts in this sense were done in the past \cite{Jacobson:2003bn,Mitro} (but later on the relevant observation~\cite{CB} appeared controversial) but so far we do not have sources for which the polarization is detected and the spectral redshift is precisely determined. In \cite{Stecker:2011ps} this problem was circumvented by using indirect methods (the same used to use GRBs as standard candles) for the estimate of the redshift. This leads to a possibly less robust but striking constraints $|\xi^{(3)}| \lesssim 2.4\times10^{-14}$. 

Remarkably this constraint was recently further improved by using the INTEGRAL/IBIS observation of the GRB 041219A, for which a luminosity distance of 85 Mpc ($z\approx 0.02$) was derived thanks to the determination of the GRB's host galaxy. In this case a constraint $|\xi^{(3)}| \lesssim 1.1\times10^{-14}$ was derived \cite{Laurent:2011he}.\footnote{The same paper claims also a strong constraint on the parameter $\xi^{(4)}$. Unfortunately, such a claim is based on the erroneous assumption that the EFT order six operators responsible for this term imply opposite signs for opposite helicities of the photon. We have instead seen that the CPT evenness of the relevant dimension six operators imply a helicity independent dispersion relation for the photon (see eq.\eqref{eq:disp-rel-dimsix}).}

%Recently, a claim of $|\xi^{(3)}| \lesssim 2 \times 10^{-7}$ was made using UV/optical polarisation measures from GRBs \cite{Fan:2007zb}. However, the strongest constraint to date comes from a local object. In \cite{Maccione:2008tq} the constraint $|\xi^{(3)}| \lesssim 6 \times 10^{-10}$ at 95\% Confidence Level (CL) was obtained by considering the observed polarization of hard-X rays from the Crab Nebula (CN) \cite{integralpol} (see also \cite{Forot:2008ud}).
%This is currently the strongest constraint on the LV coefficient for photons in the modified QED considered here.

\subsubsection{Synchrotron radiation} 
\label{synchrotron}
Synchrotron emission is strongly affected by LV, however for Planck scale LV and observed energies, it is a relevant ``window'' only for CPT odd dimension five LV QED (and dimension four LV QED, which we describe below). We shall work out here the details of CPT-odd dimension five QED ($n=3$) for illustrative reasons, for the lower dimension case see \cite{Altschul:2005za}. 

In both LI and LV cases \cite{Jacobson:2005bg}, most of the radiation from an electron of energy $E$ is emitted at a critical frequency
\begin{equation}
 \omega_c = \frac{3}{2}eB\frac{\gamma^3(E)}{E} 
% =\frac{3}{2}\frac{eB}{m_{\rm e}}\gamma^2\;,
\label{eq:omega_sync}
\end{equation}
where $\gamma(E) = (1-v^2(E))^{-1/2}$, and $v(E)$ is the electron
group velocity. 

However, in the LV case, and assuming specifically $n=3$, the electron group velocity is given by
\begin{equation}
v(E)= \frac{\partial E}{\partial p} =\left(1-\frac{m_e^2}{2p^2}+\eta^{(3)}\frac{p}{M}\right)\,.
% v(E)= \frac{\partial E}{\partial p} =\frac{p}{E}\left(1+\frac{3}{2}\eta^{(3)}\frac{p}{M}\right)\,.
 % \simeq 1-\frac{m_e^2}{2E}+\eta\frac{E}{M} + O(\eta^2)\;.
\end{equation}
Therefore, $v(E)$ can exceed $1$ if $\eta > 0$ or it can be strictly less
than $1$ if $\eta < 0$. %, resulting in $\gamma(E) \lessgtr E/m_e$ for $\eta \lessgtr 0$. 
This introduces a fundamental difference between particles with positive or negative
LV coefficient $\eta$. 

If $\eta$ is negative the group velocity of the electrons is strictly less than the (low energy) speed of light. This implies that, at sufficiently high energy, $\gamma(E)_{-} < E/m_e$, for all $E$. 
As a consequence, the critical frequency $\omega_c^{-}(\gamma, E)$ is always less than a maximal frequency $\omega_c^{\rm max}$. Then, if synchrotron emission up to some maximal frequency $\omega_{\rm obs}$ is observed, one can deduce that the LV coefficient for the corresponding leptons cannot be more negative than the value for which $\omega_c^{\rm max}=\omega_{\rm obs}$, leading to the bound~\cite{Jacobson:2005bg}
\begin{equation}
\eta^{(3)}>-\frac{M}{m_e}\left(\frac{0.34\, eB}{m_e\,\omega_{\rm obs}}\right)^{3/2}\;.
\end{equation}

If $\eta$ is instead positive the leptons can be superluminal. One can show that at energies $E_c \gtrsim 8~\TeV /\eta^{1/3}$, $\gamma(E)$ begins to increase faster than $E/m_e$ and reaches infinity at a finite energy, which corresponds to the threshold for soft VC emission. The critical frequency is thus larger than the LI one and the spectrum shows a characteristic bump due to the enhanced $\omega_c$. 

How the synchrotron emission processes at work in the CN would appear in a ``LV world'' has been studied in \cite{Jacobson:2002ye,Maccione:2007yc}. There the role of LV in modifying the characteristics of the Fermi mechanism (which is thought to be responsible for the formation of the spectrum of energetic electrons in the CN \cite{Kirk:2007tn}) and the contributions of vacuum \v{C}erenkov and helicity decay were investigated  for $n=3$ LV. This procedure requires fixing most of the model parameters using radio to soft X-rays observations, which are basically unaffected by LV.

\begin{figure}[tbp]
 \sidecaption
% \centering
 \includegraphics[scale = 0.3, angle = 90]{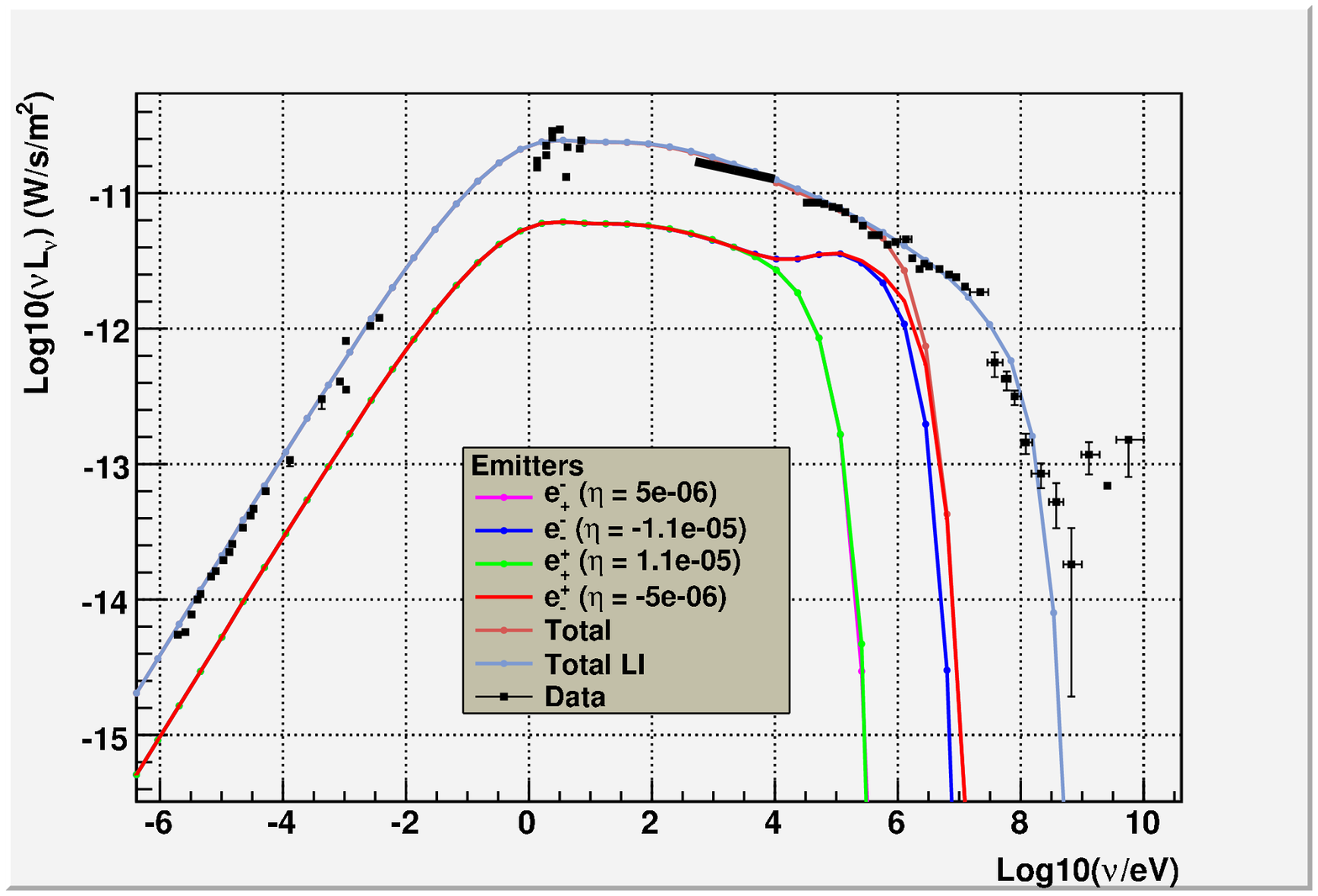}
 \includegraphics[scale = 0.3, angle = 90]{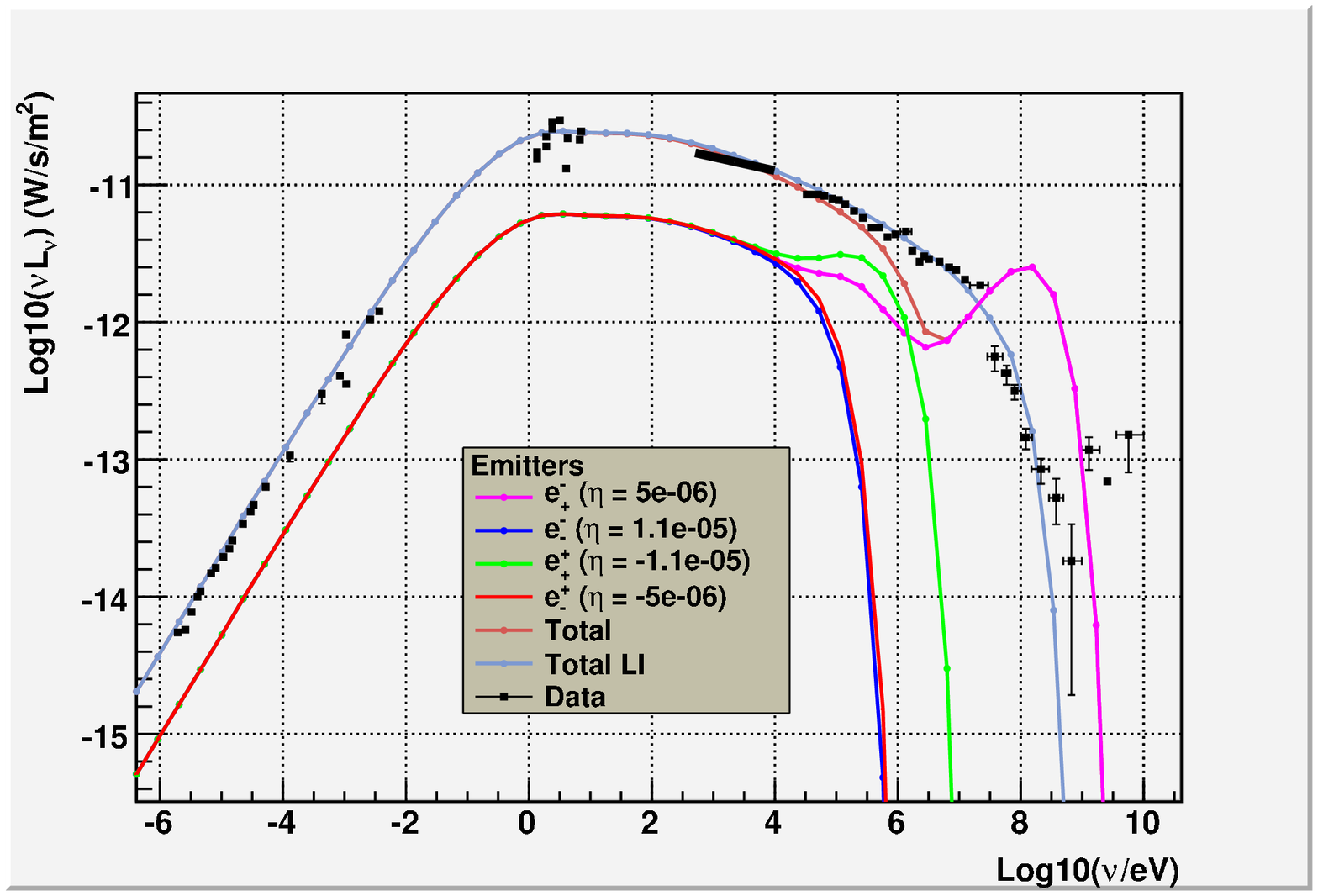}
 \caption{Comparison between observational data, the LI model and a LV
 one with $\eta_+\cdot\eta_- <0$ (left) and $\eta_+\cdot\eta_- >0$
 (right). The values of the LV coefficients, reported in the
 insets, show the salient features
 of the LV modified spectra. The leptons are injected according to the
 best fit values $p=2.4$, $E_c=2.5$ PeV. The individual contribution
 of each lepton population is shown. Picture taken from \cite{Maccione:2007yc}.}
 \label{fig:spectra}
\end{figure}

Given the dispersion relations \eqref{eq:disp_rel_phot} and \eqref{eq:disp_rel_ferm}, clearly only two configurations in the LV parameter space are truly different: $\eta_+\cdot\eta_- >0$ and $\eta_+\cdot\eta_- <0$, where $\eta_+$ is assumed to be positive for definiteness. The configuration wherein both $\eta_\pm$ are negative is the same as the $(\eta_+\cdot\eta_- >0,\,\eta_+>0)$ case, whereas that whose signs are scrambled is equivalent to the case $(\eta_+\cdot\eta_- <0,\,\eta_+>0)$. This is because positron coefficients are related to electron coefficients through $\eta^{af}_\pm = -\eta^{f}_\mp$ \cite{Jacobson:2005bg}. Examples of spectra obtained for the two different cases are shown in Fig.~\ref{fig:spectra}.

A $\chi^2$ analysis has been performed to quantify the agreement between models and data \cite{Maccione:2007yc}.
From this analysis, one can conclude that the LV parameters for the leptons are both constrained, at 95\% CL, to be $|\eta_\pm| < 10^{-5}$, as shown by the red vertical lines in Fig.~\ref{fig:total-n3}. 
Although the best fit model is not the LI one, a careful statistical analysis (performed with present-day data) shows that it is statistically indistinguishable from the LI model at 95\% CL \cite{Maccione:2007yc}.

\subsubsection{Constraints from GZK secondaries}
The same reasoning that established constraints on the CPT even higher dimension sector from GZK secondaries can also be applied to further strengthen the available constraints on CPT odd dimension 5 LV QED. In this case the absence of relevant UHE photon flux strengthens by (at most) two orders of magnitude the constraint on the photon coefficient $\xi^{(3)}$ while an eventual detection of the expected flux of UHE photons would constrain $\eta^{(3)}$ for the electron/positron at the level of $|\eta^{(3)}|\lesssim 10^{-16}$ (see \cite{Maccione:2008iw, Liberati:2009pf} for further details) by limiting the gamma decay process.  Note however, that in this case, unlike the CPT even case, one cannot exclude that only one photon helicity survives and hence a detailed flux reconstruction would be needed.

\subsubsection{Summary of constraints}

\begin{figure}[tbp]
\sidecaption[t]
\includegraphics[scale=0.30]{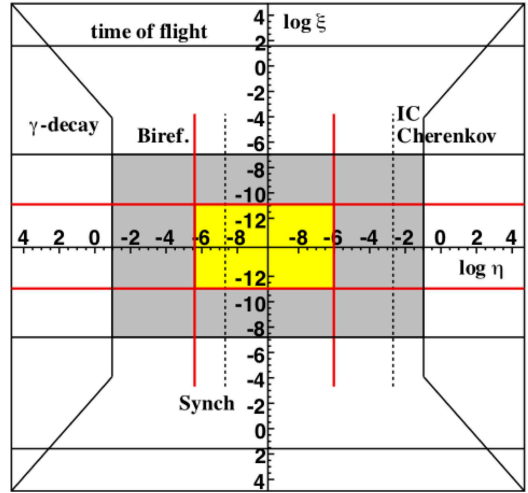} % requires the graphicx package
\caption{Summary of the constraints on LV QED at order $n=3$. The red lines are related to the constraints derived from the detection of polarized synchrotron radiation from the Crab nebula as discussed in the text. For further reference are also shown the constraints that can be derived from the detection of 80 TeV photons from the Crab nebula: the solid black lines symmetric w.r.t. the $\xi$ axis are derived from the absence of gamma decay, the dashed vertical line cutting the $\eta$ axis at about $10^{-3}$ refers to the limit on the vacuum \v{C}erenkov effect coming from the inferred 80 TeV inverse Compton electrons. The dashed vertical line on the negative side of the $\eta$ axis is showing the first synchrotron based constraint derived in~\cite{Jacobson:2002ye}.}
\label{fig:total-n3}
\end{figure}

Constraints on LV QED at $n=3$ are summarized in Fig.\ref{fig:total-n3} where also the constraints --- coming from the observations of up to 80 TeV gamma rays from the crab nebula \cite{Aharonian:2004gb} (which imply no gamma decay for these photons neither vacuum \v{C}erenkov at least up to 80 TeV for the electrons producing them via inverse Compton scattering) --- are plotted for completeness.

\subsection{Other threshold processes}

Once one realizes the power of the vacuum \v{C}erenkov effect it's natural to explore other threshold processes that might yield useful constraints.  The constraints in Figure ~\ref{fig:total-n3} are the best available constraints.  We detail below some other processes that have been used to set constraints in the past.

\subsubsection{Helicity decay}
A slightly different version of the vacuum \v{C}erenkov process is that of helicity decay (HD, $e_{\mp}\rightarrow e_{\pm}\gamma$). If $\eta_{+} \neq \eta_{-}$, an electron can flip its helicity by emitting a suitably polarized photon. This reaction does not have a real threshold, but rather an effective one \cite{Jacobson:2005bg} --- $ p_{\rm HD} = (m_e^2M/\Delta\eta)^{1/3}$, where $\Delta\eta = |\eta_+^{(3)}-\eta_-^{(3)}|$ --- at which the decay lifetime $\tau_{HD}$ changes how it scales with $\Delta\eta$. For $\Delta\eta\approx O(1)$ this effective threshold is around 10 TeV.  
Below threshold the lifetime, $\tau_{\rm HD}$ is given by $\tau_{\rm HD} > \Delta\eta^{-3} (p/10~\TeV)^{-8}\, 10^{-9}\s$, while above threshold $\tau_{\rm HD}$ becomes independent of $\Delta\eta$~\cite{Jacobson:2005bg} and is given by $\tau_{\rm HD} \approx 10^{-10} \times (10 ~{\rm Tev})/E ~\s$.   It is more difficult, however, to use helicity decay to set constraints as we do not have polarization measurements of high energy cosmic rays and the flux above a certain energy scale is only halved in the case of helicity decay, rather than almost completely removed as in the vacuum \v{C}erenkov effect.  One therefore requires more detailed knowledge of the source spectrum to properly apply helicity decay constraints. Nonetheless, HD can play a crucial role in lepton/anti-lepton propagation by basically leaving only a survival helicity state for each particle type. This mechanism, for example, played a pivotal role in the reconstruction of the Crab nebula synchrotron spectrum~\cite{Maccione:2007yc}.

\subsubsection{Photon splitting and lepton pair production}

Vacuum \v{C}erenkov radiation is an effect that would ordinarily be forbidden, however one can also look for modifications of normally allowed threshold reactions that are especially relevant for high energy astrophysics.  It is rather obvious that once photon decay and vacuum \v{C}erenkov are allowed also the related reactions in which an outgoing lepton/anti-lepton pair is replaced by two or more photons, $\gamma \rightarrow 2 \gamma$ and $\gamma \rightarrow 3 \gamma$, etc.\ , or the outgoing photons is replaced by an electron-positron pair, $e^- \rightarrow e^-e^-e^+$, are also allowed.\\
\\
\textit{LV-allowed reactions: Photon splitting}\\
\\
This is forbidden for $\xi^{(n)}<0$ while it is always allowed if  $\xi^{(n)} > 0$ \cite{Jacobson:2002hd}. When allowed, the relevance of this process is simply related to its rate. The most relevant cases are $\gamma\rightarrow \gamma\gamma$ and $\gamma\rightarrow 3\gamma$, because processes with more photons in the final state are suppressed by more powers of the fine structure constant. 

The $\gamma\rightarrow\gamma\gamma$ process is forbidden in QED because of  kinematics and CP conservation. In LV EFT both the kinematics changes and CP is not necessarily a good symmetry. However, we can argue that $\gamma \rightarrow \gamma \gamma$  is suppressed by an additional power of the Planck mass with respect to $\gamma\rightarrow 3 \gamma$. In fact, in LI QED the matrix element is zero due to the exact cancellation of fermionic and anti-fermionic loops. In LV EFT this cancellation is not exact and the matrix element is expected to be proportional to at least $(\xi E/\Mpl)^{p}$, $p>0$, as it is induced by LV and must vanish in the limit $\Mpl \rightarrow \infty$. 

Therefore we have to deal only with $\gamma\rightarrow 3 \gamma$. This process has been studied in \cite{Jacobson:2002hd,Gelmini:2005gy}. In particular, in \cite{Gelmini:2005gy} it was found that, if the ``effective photon mass'' $m_{\gamma}^{2} \equiv \xi E_{\gamma}^{n}/\Mpl^{n-2} \ll m_{e}^{2}$, then the splitting lifetime of a photon is approximately $\tau^{n=3}\simeq 0.025\,\xi^{-5} f^{-1}\left(50~\TeV/E_{\gamma}\right)^{14}~\s$, where $f$ is a phase space factor of order 1. 
This rate was rather higher than the one obtained via dimensional analysis in \cite{Jacobson:2002hd} because, due to integration of loop factors, additional dimensionless contributions proportional to $m_{e}^{8}$ enhance the splitting rate at low energy. 

This analysis, however, does not apply for the most interesting case of ultra high energy photons around $10^{19}$ eV given that at these energies $m_{\gamma}^{2} \gg m_{e}^{2}$ if $\xi^{(3)} > 10^{-17}$ and $\xi^{(4)} > 10^{-8}$. Hence the above mentioned loop contributions  are at most logarithmic, as the momentum circulating in the fermionic loop is much larger than $m_{e}$. Moreover, in this regime the splitting rate depends only on $m_{\gamma}$, the only energy scale present in the problem.
One then expects the analysis proposed in \cite{Jacobson:2002hd} to be correct and the splitting time scale to be negligible at $E_{\gamma} \simeq 10^{19}~\eV$, which therefore makes it not particularly competitive with other constraints.\\
\\
\textit{Lepton pair production}\\
\\
The processes $e^- \rightarrow e^-e^-e^+$ is similar to vacuum
\v{C}erenkov radiation or helicity decay with the final photon
replaced by an electron-positron pair.   Various combinations of helicities for the different fermions can be considered
individually. If we choose the particularly simple case (and the
only one we shall consider here) where all electrons have the same
helicity and the positron has the opposite helicity, then the threshold energy will depend on
only one LV parameter. In \cite{Jacobson:2002hd} the threshold for this reaction was derived for electron pair production, and it was found that the rate is a factor of $\sim 2.5$ times higher than that for soft vacuum \v{C}erenkov radiation. Therefore, since the rate for the reaction is  high as well, constraints may be imposed using just the value of the threshold energy.  \footnote{One could of course consider any lepton/anti-lepton pair as produced, for example the reaction $e^- \rightarrow e^- \nu \overline{\nu}$. While standard particle physics arguments imply that the rate will be roughly equivalent to the $e^-e^+$ pair production case~\cite{Ward:2012fy} given the same order of coefficients, and the threshold will be slightly lower, the constraints are on a higher dimensional parameter space and so less useful.}

\section{Dynamical frameworks III: Rotationally Invariant Minimal Standard Model Extension in QED}

\subsection{The model}
The Minimal Standard Model Extension or mSME, is the set of renormalizable operators that generate LV but maintain the existing particle content of the standard model and do not violate gauge invariance.  
A subset of the mSME, the rotationally invariant LV operators are
\begin{equation} \label{eq:LVQEDelectronrotinv}
-a u_\mu \overline{\psi} \gamma^{\mu}\psi -bu_{\mu}\overline{\psi} \gamma_5 \gamma^{\mu}\psi + \frac {1} {2} i c
u_\mu u_\nu \overline{\psi} \gamma^{\mu}  \stackrel{\leftrightarrow}{D^{\nu}} \psi  + \frac {1} {2} i
d u_\mu u_\nu \overline{\psi} \gamma_5 \gamma^\mu  \stackrel{\leftrightarrow}{D^{\nu}} \psi
\end{equation}
for fermions and
\begin{equation}\label{eq:LVQEDphotonrotinv}
-\frac{1}{4}(k_F){u_\kappa \eta_{\lambda\mu} u_\nu} F^{\kappa\lambda}F^{\mu\nu}+ \frac {1} {4} k_{AF} u^\kappa \epsilon_{\kappa \alpha \beta \gamma} A^\alpha F^{\beta \gamma}
\end{equation}
for photons.  The dimension 3, CPT odd $k_{AF}$ term generates an instability in the theory and so we will set it to zero from here on out.  The $a$ term can be absorbed by shifting the phase of the fermion field and so we will ignore it temporarily (we shall return to this term when we deal with gravity).  Note however that while the $a$ term can be absorbed in QED, it can be measured in matter sectors where the phase of the fermion is important.  For example, neutrinos can constrain differences in $a$ between species~\cite{Kostelecky:2011gq}.

The corresponding high energy ($\Mpl\gg E \gg m$) dispersion relations for QED can be expressed as (see \cite{Mattingly:2005re} and references therein for more details)
\begin{eqnarray} \label{eq:SMErotinvdisp}
E_{\rm el}^2=m_{e}^2+p^2+f^{(1)}_e p+f^{(2)}_ep^2 \quad\mbox{electrons}\\
E_{\gamma}^2=(1+ f^{(2)}_\gamma ){p^2}\quad\mbox{photons}
\end{eqnarray}
where $f^{(1)}_e=-2bs,f^{(2)}_e=-(c-ds)$, and $f^{(2)}_\gamma=k_F/2$ with $s=\pm1$ the helicity state of the fermion~\cite{Mattingly:2005re}.
The anti-fermion dispersion relation is the same as (\ref{eq:SMErotinvdisp}) with the replacement $p\rightarrow -p$, which will change only the $f^{(1)}_e$ term.

Note that the typical energy at which new phenomenology should start to appear is quite low. In fact, taking for example $f_{e}^{(2)} \sim O(1)$, one finds that the corresponding extra-term is comparable to the particle mass $m$ precisely at $p \simeq m$. Even worse, for the linear modification to the dispersion relation, we would have, in the case in which $f^{(1)}_{e} \simeq O(1)$, that $p_{\rm th} \sim m^{2}/\Mpl \sim 10^{-17}~\eV$ for electrons. (Notice that this energy corresponds by chance to the present upper limit on the photon mass, $m_{\gamma}\lesssim 10^{-18}~\eV$ \cite{pdg}.) 

\subsection{Constraints}

In contrast to the higher dimension operators, LV due to the mSME does not grow with energy.  Therefore astrophysics with its higher energies does not necessarily provide a tremendous advantage over laboratory and terrestrial experiments when testing the mSME. We list below the best constraints currently available on the QED sector of the mSME.  For a recent listing of all the constraints as well as the non-rotationally invariant case, see~\cite{Kostelecky:2008ts}.

\subsubsection{Spin polarized torsion constraints on $b$ for electrons}
Both spin polarized and unpolarized torsion balances can place limits on the mSME. Spin polarized torsion balances place limits on the electron sector of the mSME~\cite{Heckel:2008hw}, while unpolarized torsion balances constrain the gravitational sector~\cite{Kostelecky:2010ze}.  Spin polarized torsion balances constrain the electron sector, as the torsion balances are constructed to have a large number of aligned electron spins (for example, a simple magnet attached on a torsion fiber is a very crude spin polarized torsion balance).  Usually of course, the magnet design is optimized to search for LV.  For example, a vertical stack of an octagonally symmetric pattern of magnets constructed to have an overall spin polarization in the octagon's plane has been used as a torsion balance with a net electron spin polarization of $10^{23}$ electron spins~\cite{Heckel:2008hw}.  The mSME coefficients give rise to an interaction potential for non-relativistic electrons which produces an orientation dependent torque on the torsion balance which is measured using the twist of the torsion fiber.  The torsion balance in its sealed vacuum chamber is mounted on a rotating turntable, which allows for very sensitive detection of any anomalous torque as a function of the rotation frequency (and the earth's rotation and motion in the solar system). Current limits on $b$ from torsion balances are of the order of $10^{-27}$ GeV~\cite{Heckel:2008hw}.

\subsubsection{Accelerator bounds on $c$ for electrons}
Accelerator beams  of various subatomic particles are produced with (roughly) time independent energies.  Energy loss mechanisms such as the vacuum \v{C}erenkov effect $q \rightarrow q + \gamma$ (see section \ref{sec:thresholds}) for charged fermions are incompatible with a constant beam energy as long as the energy loss rate is high enough.  Limits can therefore be set on the $c$ coefficient for accelerated charged particles, as if $c$ is too much larger than zero then above the $n=2$ threshold energy \eqref{eq:energythreshvc} radiative energy losses become significant over the beam time.  The LEP experiment accelerates electrons and positrons to energies of roughly 100 GeV in the lab frame.  The beam energy is carefully measured and it is known that the beam does not lose any significant energy to vacuum \v{C}erenkov radiation.  Additionally the synchrotron emission from the beam has been measured.  Since synchrotron emission is sensitive to the Lorentz factor of the accelerated particles, the existence of synchrotron radiation of a certain spectrum given a source particle energy and path provides constraints on any LV present (see section \ref{synchrotron}).  Using the characteristics of the LEP beam and the associated synchrotron radiation spectrum limits $c$ for electrons to be of order $|c|<10^{-15}$.

\subsubsection{Astrophysical inverse Compton bounds on $d$ for electrons}
$d$ causes a spin dependent change to the dispersion relation for a fermion.  As such, one of the helicity states for a fermion is always subluminal and one can't trivially apply simple astrophysical arguments based on the vacuum \v{C}erenkov effect, etc. without knowing the helicity of the measured fermion.  However, there is a way around this by considering the dynamics of sources of high energy photons produced by inverse Compton scattering.  We receive energies from the radio up to 80 TeV from astrophysical sources such as the Crab nebula~\cite{Aharonian:2004gb}.  The overall spectrum of these sources is well understood and the high energy emission is dominated by inverse Compton scattering of accelerated electrons off of lower energy photons.  In order to constrain $d$ one can't simply require the absence of vacuum \v{C}erenkov radiation for the electrons, as half of any population will still be present under the influence of a $d$ term and available to inverse Compton scatter.  However, the magnetic field present in the same sources causes the electrons to precess, thereby destroying any initial polarization of the electrons.  Therefore, as argued in~\cite{Altschul:2006he} both helicities of electrons must be present and stable if there is to be inverse Compton radiation.  The vacuum \v{C}erenkov effect is therefore forbidden for both helicities, which allows one to put double sided constraints on $d$ of the order of $|d|<O(10^{-12})$~\cite{Altschul:2006he}.\footnote{Note that the bounds presented here are weaker by a factor of $10^{-3}$, as we have used the CMB frame as the rest frame rather than the sun centered frame and therefore the strength of the bounds are weakened by the $v/c$ of the Sun with respect to the CMB.}

\subsubsection{Cosmic ray and HESS bounds on $k_F$}
Bounds on the fermion sector that use processes involving the fundamental QED vertex implicitly assume that the photon sector is unmodified.  Which sector is the ``unmodified'' sector is arbitrary, in that the limiting speed of one of the sectors can be defined to be the ``speed of light''.  When constraining photon coefficients, we simply make the opposite assumption - we assume the fermion sector is unmodified and constrain $k_F$.  Constraints on $k_F$ can be generated for protons and photons using the same effects as above, i.e. the necessary absence of vacuum \v{C}erenkov and photon decay if we see high energy cosmic ray protons and TeV gamma rays.  In~\cite{Klinkhamer:2008ky} the authors used the necessary absence of $p\rightarrow \p + \gamma$ and a Pierre Auger event in conjuction with the excess of high energy TeV gamma rays observed by the HESS telescope (which forbids $\gamma \rightarrow p+\overline{p}$) to produce a two sided bound on $k_F$ of $-9 \times 10^{-16} < k_f < 6 \times 10^{-20}$.

\section{Dynamical Frameworks IV: Gravity}

\subsection{The model}
The SME is constructed by coupling matter terms to non-zero LV tensors in vacuum.  If we left the tensors as constants, without any sort of dynamics, then one would break general covariance.  It may be that for other reasons one may want to change the underlying symmetry structure for gravity in just this way; such is the case with Ho\v rava--Lifshitz  gravity~\cite{Horava:2009uw}.  Alternatively, if one wants to preserve general covariance one can by promoting any LV tensors to dynamical fields.  Dynamical Lorentz breaking is also sometimes called ``spontaneous Lorentz violation'', although this is a bit of a misnomer as there are models where there really is no Lorentz invariant phase.  There are many ways that a dynamical field can generate a LV term in vacuum.  If we restrict ourselves to rotational invariance, then it is natural to generate LV couplings by including in the action either a scalar or a timelike vector field that takes a vacuum expectation value.  In the case of a scalar, one can use a shift symmetry ($\phi(x) \rightarrow \phi(x) + \phi_0$) to construct actions for which the derivative of the scalar takes a non-zero value (c.f. ~\cite{ArkaniHamed:2003uy}).  In the vector case, one simply puts a potential for the vector field such that the vector acquires a vev.  We concentrate on the vector case here as it is the simplest model that allows for rotationally invariant Lorentz violation~\cite{Jacobson:2000xp}.  Just as the SME is a derivative expansion in derivatives of matter fields, one can treat the vector field in the same way.  If we still denote the vector field by $u^\alpha$ then we can write the low energy action for gravity and the vector as  

	\beql{ac:ae}
	\ac = \acEH + \ac_{u} = \fr{1}{16\pi\Gae}\intdx{4}\rt{-\met}\; \left(R + \lag_{u}\right).
	\eeq
where $\lag_{u}$ is given by
	\beql{lag:ae}
	\lag_{u} = -\tn{Z}{^{\alpha \beta}_{\gamma \delta}}(\Dl_\alpha u^\gamma)(\Dl_\beta u^\delta) + V(u).
	\eeq

The tensor $\tn{Z}{^{\alpha \beta}_{\gamma \delta}}$ is defined as~\cite{Jacobson:2008aj}\ft{Note the indicial symmetry $\tn{Z}{^{\beta \alpha}_{\delta \gamma}} = \tn{Z}{^{\alpha \beta}_{\gamma \delta}}$.}

	\beql{def:Zabcd}
	\tn{Z}{^{\alpha \beta}_{\gamma \delta}} = c_1\met^{\alpha \beta}\met_{\gamma \delta} + c_2\tn{\de}{^\alpha_\gamma}\tn{\de}{^\beta_\delta} + c_3\tn{\de}{^\alpha_\delta}\tn{\de}{^\beta_\gamma} - c_4u^\alpha u^\beta\met_{\gamma \delta}~,
	\eeq
where $c_i,\,i = 1,\ldots,4$ are simple coefficients of the various kinetic terms and $V(u)$ is a potential term that generates a non-zero vev for $u^\alpha$.  An additional term, $R_{ab} u^a u^b$ is a combination of the above terms when integrated by parts, and hence is not explicitly included here.

In general, such a theory possesses four additional degrees of freedom.  One of the vector components, however, will necessarily have a wrong sign for its kinetic term, thereby generating a ghost excitation.  This can be remedied, at least at low energies, by choosing the potential $V(u)=\lambda(u^2+1)$, where $\lambda$ is a Lagrange multiplier.  This fixes the norm of $u^\alpha$ and removes the ghost excitation \footnote{Note that one could have fixed the norm to be larger or smaller than one, however one can simply scale $u^\alpha$ to have norm $-1$ and absorb the scaling into the coefficients.}  so the excitations of the vector all have positive norm (c.f. the discussions in \cite{Eling:2004dk} and \cite{Bluhm:2008yt}).  With this form of the potential, the vector theory is known as ``Einstein-aether theory''~\cite{Jacobson:2000xp}.  We will refer to the vector field as an ``aether'' field, vector field models that couple to the SM Lagrangian are also sometime nicknamed ``bumblebee'' models.  In reality of course the Lagrange multiplier is likely simply an approximation to a potential that generates dynamics such that the ghost only has UV effects.  For the purposes of constraints, we will treat the Lagrange multiplier as ``the'' potential term and so neglect any possible ghost excitation.

The matter couplings between aether and the standard model field content are the same as they were before, only now the field has dynamics so there can be position dependent violations of Lorentz symmetry.  Interestingly, some of the couplings to matter that are unobservable for a single fermion field can have relevant effects when the aehter varies.  For example, the $-a u_\mu \overline{\psi} \gamma^\mu \psi$  term in the mSME could be removed by making a phase change for the fermion.  However, once $u^\a$ is dynamical and varies with position, only a single component of the term can actually be removed by a phase change~\cite{Kostelecky:2008in}.  This leads to a new type of term which requires gravitational/position dependent tests in the matter sector.~\cite{Kostelecky:2008in}.  

Additionally, just as constraints can be put on the coupling of matter fields to the LV vector field, constraints can also be put on the $c_i$ coefficients themselves by examining the relevant PPN parameters and comparing to known PPN limits from solar system and other observations.

\subsection{Constraints}

\subsubsection{Constraints on the aether kinetic terms}

\textit{Constraints from PPN analysis}\\
\\
There are two primary methods to constrain the aether kinetic terms.  First, one can compute the ``parameterized post-Newtonian'' or PPN coefficients for the theory and compare to observational tests.  The only two non-vanishing PPN parameters are $\a_1,\a_2$ which describe preferred frame effects.  $\a_1$ and $\a_2$ have been calculated~\cite{Foster:2005dk} as
\begin{eqnarray}
\alpha_1=\frac {-8(c_3^2+c_1c_4)} {2c_1-c_1^2+c_3^2}\\
\alpha_2=\frac {\a_1} {2}- \frac {(c_1+2c_3-c_4)(2c_1+3c_2+c_3+c_4)} {(c_1+c_2+c_3)(2-c_1-c_4)}.
\end{eqnarray}
Current constraints are $\a_1<10^{-4}$ and $\a_2<4 \times 10^{-7}$~\cite{Will:2005va} and so from a PPN analysis alone there is still a large 2-d region of parameter space that remains consistent with available tests of GR.\\
\\
\textit{Constraints from gravity-aether wave modes}\\
\\
The aether introduces three new excitations into the gravity sector (there are naively four, but the unit constraint removes one).  These excitations strongly couple to the metric via the constraint.  The combined aether-metric modes consist of the two usual transverse traceless graviton modes, a vector mode, and a scalar mode~\cite{Jacobson:2004ts}.  Gravitational wave detectors can in principle test for polarizations~\cite{Chatziioannou:2012rf}, however there is an additional possibility to constrain the kinetic terms.  The speeds of each of the modes can differ from the speed of light.  Hence if the speeds are less than unity, high energy cosmic rays will emit vacuum gravitational \v{C}erenkov radiation~\cite{Elliott:2005va}.  If we denote the speeds of the spin-2, spin-1 and spin-0 modes by $s_2, s_1, s_0$ then we have~\cite{Jacobson:2004ts}
\begin{eqnarray}
s_2^2=(1-c_1-c_3)^{-1}\\
s_1^2=\frac{2c_1-c_1^2+c_3^2} {2(c_1+c_4)(1-c_1-c_3)}\\
s_0^2=\frac{(c_1+c_2+c_3)(2-c_1-c_4)}{(c_1+c_4)(1-c_1-c_3)(2+3c_2+c_1+c_3)}.
\end{eqnarray}

Requiring all these speed to be greater than unity therefore puts constraints on a combination of the $c_i$ coefficients. \\
\\
\textit{Combined constraints}\\
\\
Even after imposing the above constraints there is still a large region of parameter space allowed.  To get an estimate of the size of the space, one can set $\a_1$ and $\a_2$ equal to zero and solve the resulting equations for $c_2$ and $c_4$ in terms of $c_1$ and $c_3$.  If we define $c_+=c_1+c_3$ and $c_-=c_1-c_3$ then the PPN and gravitational \v{C}erenkov constraints are all satsified provided~\cite{Jacobson:2008aj}
\begin{eqnarray}
0\leq c_+\leq 1\\
0\leq c_- \leq \frac {c_+} {3(1-c_+)}
\end{eqnarray}
which shows that the gravitational sector is only minimally constrained compared to aether-matter couplings.

\subsubsection{Gravitational constraints on aether-matter couplings}

There are couplings between aether and matter that can only be strongly tested when the aether is dynamical.  The prime example is the coupling $-a u_\mu \overline{\psi} \gamma^{\mu}\psi$ in the mSME.  As said, when gravity is neglected and we're dealing with an observation that involves only one fermion field, then the $a u_\mu$ term can be absorbed by making a phase transformation of the fermion $\psi\rightarrow e^{i a u \cdot x} \psi$.  This cannot be done when $u$ is a function of position, which would necessarily be the case if $u$ is a dynamical field strongly coupled to gravity (as we've done with the constraint above).  One can therefore use gravitational experiments to measure the position dependence of $u^\mu$, similar to how redshift experiments measure the gravitational field.  For a larger overview of constraints, see ~\cite{Kostelecky:2008ts, Hohensee:2011wt}.  We will concentrate here on a recently applied method, that of atom interferometry~\cite{Hohensee:2011wt}.

Atom interferometry involves splitting of a beam of cold atoms such as {\it Cs} along two beam arms and comparing the phase difference at a point of the Cs atoms that travel along each beam.  Since gravity and the matter sector of the mSME affect the phase of the atomic wavefunctions, one can look for an anomalous sidereal pattern.  By orienting the interferometer vertically and horizontally, one can adjust the sensitivity to the gravitational and matter sectors of the mSME.  Since one can also use other limits to control for any possible matter mSME effect on the signal one can thereby isolate solely gravitational effects.  Hence atom interferometers have been used with great success on constraining the gravitational sector of the mSME.  

Consider an interferometer that is constructed to measure phase shifts of of atoms moving vertically vs. horizontally.  There will be a contribution to the total phase shift $\Delta \phi$ that comes from the gravitational redshift, i.e. some $\Delta \phi_{red}$.  In the Newtonian limit where $u_\mu$ is proportional to the timelike Killing vector, $u^\mu$ is given by $u^\mu=(1-U,0,0,0)$ ~\cite{Carroll:2004ai}, where $U$ is the Newtonian gravitational potential.  If we couple $u^\mu$ to matter, one can easily see that the effect of the coupling via an $a u^\mu$ or $c u^\mu u^\nu$ term is to change the redshift by an additional amount proportional to the gravitational potential, i.e. the change in frequency of a wave is given by $\Delta \omega = (1+\beta) \Delta U$ where $\beta$ is an experimental parameter related to the fundamental Lagrangian parameters (c.f. ~\cite{Muller:2010zzb}) for a discussion.  The phase shift generated by this extra gravitational redshift is then just $\Delta \phi_{red}=(1+\beta) \Delta \phi_0$, where $\Delta \phi_0$ is the expected gravitational redshift.  Constraints on $\beta$ therefore limit the fundamental Lagrangian parameters.  $\beta$ is limited to be zero within a few parts per billion, which then generates fairly strong limits on position dependence of the $a$ and $c$ couplings~\cite{Hohensee:2011wt}.

\section{Neutrinos}
While we have concentrated primarily on QED, we would be remiss to not mention at least in passing the important role of neutrinos in tests of Lorentz violation.  More so than other particles, neutrinos are uniquely suited to test various aspects of Lorentz symmetry.  They are copiously produced in both terrestrial experiments and astrophysical objects, so there are solar neutrinos at $<1$ MeV, controlled beams of neutrinos at roughly 10 GeV, atmospheric neutrinos up to 10 TeV, and (theoretically) ultra-high energy neutrinos up to $10^{18}$ eV and above.   Neutrinos are weakly interacting, so they propagate over long distances which allows for detailed time of flight and threshold reaction analyses.  Neutrinos have a very small mass, which make threshold analyses even more sensitive.  Finally, neutrinos oscillate between flavor eigenstates, which constrains interspecies Lorentz violation.  

The problem, of course, is that neutrinos are rather difficult to detect.  However, thanks to the numerous ongoing neutrino experiments, many significant experimental results have been published that can be adapted to constraining Lorentz violation in the neutrino sector.  We now turn to the theoretical framework and current and future constraints on the neutrino sector.

\subsection{Neutrinos I: Species dependent Lorentz violation}
\subsubsection{The model}
Neutrinos come in three distinct flavors/masses, and there is \textit{a priori} no reason to believe that any Lorentz violating coefficients are the same for each species.  It may even be that any Lorentz violation is not diagonal in either basis.  We make a couple simplifying assumptions on this point - that the neutrinos are Dirac neutrinos and that any Lorentz violation is diagonal in the mass basis.  This is a reasonable starting point, as one natural idea is that since any theory of quantum gravity must reduce to general relativity in the infrared, any Lorentz violation induced by quantum gravity would be primarily controlled by the charges that couple to gravity.  However, this does not meant hat the coefficients for each mass eigenstate is the same.  Indeed, one would expect that, due to RG effects, even if the coefficients were the same for all eigenstates at one energy they would not be the same for the large range of energies we can test in neutrino experiments.  With these assumptions, the Lorentz violating terms (written in the mass basis) are exactly those for the QED fermions,
\begin{eqnarray}
-a_i u_\mu \overline{\psi}_i \gamma^{\mu}\psi_i -b_i u_{\mu}\overline{\psi}_i \gamma_5 \gamma^{\mu}\psi_i + \frac {1} {2} i c_i u_\mu u_\nu \overline{\psi}_i \gamma^{\mu}  \stackrel{\leftrightarrow}{\partial^{\nu}} \psi_i  \\ \nonumber + \frac {1} {2} i
d_i u_\mu u_\nu \overline{\psi}_i \gamma_5 \gamma^\mu  \stackrel{\leftrightarrow}{\partial^{\nu}} \psi_i +\frac{1}{2\Mpl}u^m\overline{\psi}_i\gamma_m(\zeta_{i,1}+\zeta_{i,2}\gamma_5)(u\cdot\partial)^2\psi_i\\ \nonumber
 - \frac{i}{\Mpl^{2}}\overline{\psi}_i(u\cdot \partial)^{3}(u\cdot \gamma)(\alpha_{i,L}^{(6)}P_{L} + \alpha_{i,R}^{(6)}P_{R}) \psi_i
\end{eqnarray}
where $i$ is the mass index.  We've dropped the gauge covariant derivative above, as it is irrelevant and couples in the flavor basis so merely would add needless complication.   Also note that we have included both right and left projection operators.  However, since standard model interactions only produce left-handed neutrinos, the constraints will primarily be on the corresponding left-handed operators and so we drop all $P_R$ terms in our discussion of oscillations (although we will need to return to them when we discuss neutrino splitting).  Finally, in contrast to the QED case, we cannot drop the $a_i$ term as this gives a contribution to the oscillation pattern.

The above terms and the usual Dirac Lagrangian for the neutrino yield a high energy neutrino dispersion relation of
\begin{eqnarray}
\label{eq:disp_rel_nu}
E_i^2 = p^2 + N_i^2 \\
\nonumber N_i^2=m_i^2 + 2(a_i+b_i)p-(c_i+d_i) p^2+ \frac{m_i}{\Mpl}\alpha_{i,L}^{(5)}p^{2}\\ \nonumber
 + 2(\zeta_{i,1}-\zeta_{i,2}) \frac{p^3}{\Mpl} + 2 \frac{\alpha_{i,L}^{(6)}p^{4}}{\Mpl^{2}} 
\end{eqnarray}

The tightest constraints on species dependence comes from neutrino flavor oscillation measurements.  Neutrino oscillations depend on the differences in $E-p$ between different neutrino eigenstates.  In standard neutrino oscillations, this difference is governed by the squared mass differences between the energy eigenstates. With Lorentz violation and our assumption that the Lorentz violating eigenstates are the mass eigenstates oscillations are governed by the differences in the effective mass squared.  Therefore, neutrino oscillations do not probe any absolute Lorentz violation in the neutrino sector, but rather the differences in the dispersion relations between different neutrino states. 

Now let us consider a neutrino produced via a particle reaction in a definite flavor eigenstate $I$ with momentum $p$. The amplitude for this neutrino to be in a particular mass eigenstate $i$ is represented by the matrix $U_{Ii}$, where $\sum U_{Ji}^\dagger U_{Ii} = \delta_{IJ}$. The amplitude for the neutrino to be observed in another flavor eigenstate $J$ at some distance $L$ from the source, after some time $T$  is then
\begin{equation}
A_{IJ} = \sum_i U_{Ji}^\dagger e^{-i(ET-pL)} U_{Ii} \approx \sum_I U_{Ji}^\dagger e^{-i LN_i^2/(2E)}U_{Ii}\;.
\end{equation}
The transition probability can then be written as 
\begin{equation}
P_{IJ} = \delta_{IJ} - \sum_{i,j>i}4F_{IJij}\sin^2\left(\frac{\delta N_{ij}^2L}{4E}\right)+2G_{IJij}\sin^2\left(\frac{\delta N_{ij}^2L}{2E}\right)\;,
\end{equation}
with $\delta N_{ij}^2 = N_i^2-N_j^2$ and $F_{IJij}$ and $G_{IJij}$ are functions of the mixing matrixes.  With this formalism one can compare with existing neutrino oscillation experiments that measure $P_{IJ}$.  Many of these experiments quote results on a deviation of the neutrino speed from that of light, i.e.
\begin{equation}
\left(\frac{\Delta c}{c}\right)^{LIV}_{ij}= E^{-2}(\delta N_{ij}^2 - \delta m_{ij}^2)
\end{equation}
which can be easily translated into a constraint on the coefficients in the Lagrangian, as we do below.
\subsubsection{Constraints}
We now turn to the constraints that can be put on the neutrino sector.  For the renormalizable terms governed by $a_i,b_i,c_i,d_i$ and the non-renormalizable $\alpha_{i,L}^{(5)}$ term that contributes a similar term in the dispersion relation the best known limits come from the miniBoone experiment~\cite{AguilarArevalo:2011yi}. They limit various combinations of coefficients at the $10^{-20}$ GeV level, which implies that the order of magnitude constraints (assuming no cancellations) on these coefficients are
\begin{eqnarray}
|a_i|,|b_i| \lesssim 10^{-20} \mathrm{GeV}\\ \nonumber
|c_i|,|d_i| \lesssim 10^{-20} \\ \nonumber
\alpha_{i,L}^{(5)} \lesssim 10^9
\end{eqnarray}
where we have used the mass of the neutrino as approximately 0.1 eV and the energy of the neutrinos in the miniBoone experiment as approximately 1 GeV (it's actually slightly less). As one can see, both of the renormalizable operators are very tightly constrained, while the non-renormalizable operator is essentially free.

For the higher dimension operators that contribute higher order corrections to the dispersion relation the best constraints to date come from the survival of atmospheric muon neutrinos observed by the former IceCube detector AMANDA-II in the energy range 100 GeV to 10 TeV \cite{Kelley:2009zza}.  AMANDA-II searched for a generic LIV in the neutrino sector~\cite{GonzalezGarcia:2004wg} and achieved $(\Delta c /c)_{ij} \leq 2.8\times10^{-27}$ at 90\% confidence level assuming maximal mixing for some of the combinations $i,j$. Using the low end of the energy band (100 GeV) to be conservative, this yields order of magnitude constraints on the Lorentz violating coefficients of
\begin{eqnarray}
 |\zeta_{i,1}|,|\zeta_{i,2}|\lesssim 10^{-10}\\ \nonumber 
 |\alpha_{i,L}^{(6)}| \lesssim 10^7.
\end{eqnarray}
Given that IceCube does not distinguish neutrinos from antineutrinos, the same constraints apply to the corresponding antiparticles. Of interest is that due to the strong energy dependence of the dimension six term, as more data is taken at the 10 TeV range the constraint will drop below $O(1)$ (at 10 TeV for the energy the constraint is already $\alpha_{i,L}^{(6)}<0.1$).
The IceCube detector is expected to improve this constraint to $(\Delta c / c)_{ij} \leq 9\times 10^{-28}$ in the next few years \cite{Huelsnitz:2009zz}. We also note that the lack of sidereal variations in the atmospheric neutrino flux also yields comparable constraints on some combinations of SME parameters \cite{Abbasi:2010kx}, which can be translated into the framework above.
Finally, a nice summary of neutrino oscillation observations, with particular attention to LIV, can be found in \cite{Diaz:2011ia}.  For a comprehensive listing of constraints on terms in the neutrino sector of the SME see~\cite{Kostelecky:2008ts,Kostelecky:2011gq}.

\subsection{Neutrinos II: Species independent Lorentz violation}

Neutrino oscillation depend on $\delta N_{ij}$ and so if all the Lorentz violating terms are species independent there is no contribution to the oscillation pattern from Lorentz violation.  Hence other methods must be used to constrain these terms.  The model is almost the same as above.  The only differences are that all neutrinos have the same Lorentz violating coefficients and therefore the $a_i$ term can be dropped as in QED. Hence we proceed directly to the relevant constraints.

\subsubsection{Constraints}

\textit{Time of flight}
\\
For pure time of flight constraints we have to date only two observations to rely on, the supernova SN1987a neutrino burst and the ICARUS experiment. We deal first with SN1987, which was a unique event that generated the almost simultaneous (within a few hours) arrival of electronic antineutrinos and photons. Although only few electronic antineutrinos at MeV energies was detected by the experiments KamiokaII, IMB and Baksan, it was enough to establish a constraint $(\Delta c/c)^{TOF} \lesssim 10^{-8}$ \cite{Stodolsky:1987vd} or $(\Delta c/c)^{TOF} \lesssim 2\times10^{-9}$ \cite{Longo:1987gc} by looking at the difference in arrival time between antineutrinos and optical photons over a baseline distance of $1.5\times10^5$ ly. Further analyses of the time structure of the neutrino signal strengthened this constraint down to $\sim10^{-10}$ \cite{Ellis:2008fc,Sakharov:2009sh}. 

The scarcity of the detected neutrino did not allow the reconstruction of the full energy spectrum and of its time evolution in this sense one should probably consider constraints purely based on the difference in the arrival time with respect to photons more conservative and robust. Adopting $\Delta c/c \lesssim 10^{-8}$, the SN constraint implies the following order of magnitude constraints
\begin{eqnarray}
|b| \lesssim 10^{-11} \mathrm{GeV}\\ \nonumber
|c|,|d_i| \lesssim 10^{-8} \\ \nonumber
\alpha_{L}^{(5)} \lesssim 10^21\\ \nonumber
 |\zeta_{1}|,|\zeta_{2}|\lesssim 10^{13}\\ \nonumber 
 |\alpha_{L}^{(6)}|\lesssim 10^{34}.
\end{eqnarray}
Hence time of flight constraints are quite tight for renormalizable operators but leave the non-renormalizable operators effectively unconstrained.   The ICARUS experiment measured the time of flight of neutrinos traveling from CERN to Gran Sasso in a repeat of the OPERA experiment.  ICARUS found that the arrival time was consistent with zero and within approximately 10 ns of the expected light travel time~\cite{Antonello:2012hg}.  The light travel time betwen CERN and Gran Sasso is roughly 2.4 ms, so the $\Delta c/c$ from ICARUS is of the order of $10^{-5}$, consistent with previous measurements made by the MINOS detector~\cite{Adamson:2007zzb}.  Therefore the SN1987A neutrinos remain the tightest constraint on Lorentz violation from time of flight experiments.\\
\\
\textit{Threshold reactions}\\
\\
Threshold reactions also can be used to cast constraints on the neutrino sector. Several processes are of interest: neutrino \v{C}erenkov emission $\nu\to\gamma\,\nu$, neutrino splitting $\nu\to \nu\,\nu \overline{\nu}$, and neutrino electron/positron pair production $\nu\to\nu \,e^{-}e^{+}$.  Let us consider for illustration the latter process as the others work similarly.  Neglecting possible LV modification in the electron/positron sector (on which we have seen we have already strong constraints) the threshold energy for a dispersion modification that scales as $p^n$ is
\begin{equation}
E_{th, (n)}^{2} = \frac{4m_{e}^{2}}{\delta_{(n)}}\;,
\end{equation}
with $\delta_{(n)} = \xi_{\nu}(E_{th}/M)^{n-2}$.

The rate of this reaction was firstly computed in \cite{Cohen:2011hx} for $n=2$ but can be easily generated to arbitrary $n$~\cite{Maccione:2011fr}. The generic energy loss time-scale then reads (dropping purely numerical factors)
\begin{equation}
\tau_{\nu{\rm -pair}} \simeq \frac{ m_Z^4 \cos^4 \theta_w}{g^4E^5} \left( \frac {M} {E} \right)^{3(n-2)}\;,
\end{equation}
where $g$ is the weak coupling and $\theta_w$ is Weinberg's angle.  ICARUS found no electron-positron pair creation from the CERN neutrino beam as it passed through their detector~\cite{ICARUS:2011aa}, but the best constraint comes from the observation of upward-going atmospheric neutrinos up to 400 TeV by the IceCube experiment. Since the neutrinos propagated through the entire Earth to reach the IceCube detector the free path of these particles is at least longer than the Earth radius.  This measurement has been used to establish constraints on $|\zeta_{1}|,|\zeta_{2}|< 30$ while $c,d$ are constrained at the $10^{-12}$ level. No effective constraint can be optioned for the dimension six operators, however in this case neutrino splitting (which has the further advantage to be purely dependent on LV on the neutrinos sector) could be used on the ``cosmogenic'' neutrino flux. This is supposedly created via the decay of charged pions produced by the aforementioned GZK effect. The neutrino splitting should modify the spectrum of the ultra high energy neutrinos by suppressing the flux at the highest energies and enhancing it at the lowest ones. In \cite{Mattingly:2009jf} it was shown that future experiments like ARIANNA \cite{Barwick:2006tg} will achieve the required sensitivity to cast a constraint of order $|\alpha_{L}^{(6)}| \lesssim 10^{-4}$. Note however, that the rate for neutrino splitting computed in \cite{Mattingly:2009jf} was recently recognized to be  underestimated by a factor $O(E/M)^2$~\cite{Ward:2012fy}. Hence the future constraints here mentioned should be recomputed and one should be able to strengthen the constraint by a few orders of magnitude.

\section{Other frameworks}
\label{sec:others}

Specifying which dynamical framework is employed is crucial when discussing the phenomenology of Lorentz violations. The most robust and well-motivated framework is that which we have been discussing, effective field theory.  However, it is not the only one and, in fact, there are reasonable arguments from holography that a quantum gravity theory should not necessarily be a local field theory in the UV (c.f. the discussion in ~\cite{Shomer:2007vq}).  Hence Lorentz violation may enter into low energy physics in novel ways.  In addition, if one believes that the fundamental quantum gravity theory should not be Lorentz invariant, then one might look for ways in which Lorentz violation might appear outside the realm of effective field theory and so avoid many of the constraints that exist in the EFT framework.   For completeness, and because the EFT approach is nothing more than a highly reasonable, but rather arbitrary ``assumption'', it is worth studying and constraining additional models, given that they may evade the majority of the constraints discussed in this review.

%In fact, not all the above mentioned ``windows on quantum gravity" can be exploited without adding additional information about the dynamical framework one works with. Although cumulative effects exclusively use the form of the modified dispersion relations, all the other ``windows" depend on the underlying dynamics of interacting particles and on whether or not the standard energy-momentum conservation holds. Thus, to cast most of the constraints on dispersion relations of the form (\ref{eq:disprel}), one needs to adopt a specific theoretical framework justifying  the use of such deformed dispersion relations.  

%The previous discussion mainly focuses on considerations based on Lorentz breaking EFTs. This is indeed  a conservative framework within which much can be said (e.g.~reaction rates can still be calculated) and from an analogue gravity point of view it is just the natural frame to work within.  Nonetheless, this is of course not the only dynamical framework within which a Lorentz breaking kinematics can be cast.

\subsection{D-brane models}
\label{sec:nonEFT}

Ellis et. al. developed a model  \cite{Ellis:1999jf,Ellis:2003sd} in which modified dispersion relations derive from the Liouville string approach to quantum space-time \cite{Ellis:1992eh}. Liouville-string models \cite{Ellis:1992eh} motivate corrections to the usual relativistic dispersion relations that are first order in the particle energies and that correspond to a vacuum refractive index $\eta = 1-(E/\Mpl)^{\alpha}$, where $\alpha = 1$. Models with quadratic dependences of the vacuum refractive index on energy: $\alpha = 2$ have also been considered \cite{Burgess:2002tb}.

In particular, the D-particle realization of the Liouville string approach predicts that only gauge bosons such as photons, and not charged matter particles such as electrons, might have QG-modified dispersion relations. This difference occurs since excitations which are charged under the gauge group are represented by open strings with their ends attached to the D-brane \cite{Polchinski:1996na}, and that only neutral excitations are allowed to propagate in the bulk space transverse to the brane \cite{Ellis:2003if}. Thus, if we consider photons and electrons, in this model the parameter $\eta$ is forced to be null, whereas $\xi$ is free to vary. Even more importantly, the theory is CPT even, implying that vacuum is not birefringent for photons ($\xi_{+} = \xi_{-}$).

\subsection{New Relativity Theories}

Lorentz invariance of physical laws relies on only a few assumptions: the principle of relativity, stating the equivalence of physical laws for non-accelerated observers, isotropy (no preferred direction) and homogeneity (no preferred location) of space-time, and a notion of precausality, requiring that the time ordering of co-local events in one reference frame be preserved \cite{ignatowsky,ignatowsky1,ignatowsky2,ignatowsky3,ignatowsky4,Liberati:2001sd,Sonego:2008iu,Baccetti:2011aa}. 

All the realizations of LV we have discussed so far explicitly violate the principle of relativity by introducing a preferred reference frame. This may seem a high price to pay to include QG effects in low energy physics. For this reason, it is worth exploring an alternative possibility that keeps the relativity principle but that relaxes one or more of the above postulates. 

For example, relaxing the space isotropy postulate leads to the so-called Very Special Relativity framework \cite{Cohen:2006ky}, which was later on understood to be described by a Finslerian-type geometry~\cite{Bogoslovsky:2005cs,Bogoslovsky:2005gs,gibbons:081701}. In this example, however, the generators of the new relativity group number fewer than the usual ten associated with Poincar\'e invariance. Specifically, there is an explicit breaking of the $O(3)$ group associated with rotational invariance. 

One may wonder whether there exist alternative relativity groups with the same number of generators as special relativity. Currently, we know of no such generalization in coordinate space. However, it has been suggested that, at least in momentum space, such a generalization is possible, and it was termed  ``doubly" or ``deformed" (to stress the fact that it still has 10 generators) special relativity (DSR).  Even though DSR aims at consistently including dynamics, a complete formulation capable of doing so is still missing, and present attempts face major problems. Thus, at present DSR is only a kinematic idea. 

Finally we cannot omit the recent development of what one could perhaps consider a spin-off of DSR that is Relative Locality, which is based on the idea that the invariant arena for classical physics is a curved momentum space rather than spacetime (the latter being a derived concept)~\cite{AmelinoCamelia:2011bm}.

DSR and Relative Locality are still a subject of active research and debate (see e.g.~\cite{Smolin:2008hd,Rovelli:2008cj,AmelinoCamelia:2011uk,Hossenfelder:2012vk}); nonetheless, they have not yet attained the level of maturity needed to cast robust constraints\footnote{Note however, that some knowledge of DSR phenomenology can be obtained by considering that, as in Special Relativity, any phenomenon that implies the existence of a preferred reference frame is forbidden. Thus, the detection of such a phenomenon would imply the falsification of both special and doubly-special relativity. An example of such a process is the decay of a massless particle.}. 
\section{Discussion and Perspectives}

%We can summarize the current status of the constraints for the LV SME in the following table.
%\begin{table}[!htb]
%\caption{Summary of typical strengths of the available constrains on the SME at different orders.}
% Follow this input for your own table layout
%
%\begin{center}
%\begin{tabular}{p{1.2cm}|p{2.5cm}|p{2.5cm}|p{2.5cm}|p{2.5cm}}
%\hline\noalign{\smallskip}
%Order & photon & $e^{-}/e^{+}$ & Hadrons & Neutrinos$^a$  \\
%\noalign{\smallskip}\svhline\noalign{\smallskip}
%n=2 & N.A. & $O(10^{-16})$ & $O(10^{-27})$ & $O(10^{-8})$ \\
%n=3 & $O(10^{-14})$ (GRB) & $O(10^{-16})$ (CR) 
%%$O(10^{-8})$ (CN) 
%& $O(10^{-14})$ (CR) & $O(30)$ \\
%n=4 & $O(10^{-8})$ (CR) & $O(10^{-8})$ (CR) & $O(10^{-6})$ (CR)  & $O(10^{-4})^*$ (CR) \\
%\end{tabular}
%\end{center}
%GRB=gamma rays burst, CR=cosmic rays\\
%% CN=Crab Nebula.\\
%$^a$ From neutrino oscillations we have constraints on the difference of LV coefficients of different flavors up to $O(10^{-28})$ on dim 4, $O(10^{-8})$ and expected up to $O(10^{-14})$  on dim 5 (ICE3), expected up to $O(10^{-4})$ on dim 6 op.
%$^*$ Expected constraint from future experiments.  
%\label{table:sum} 
%\end{table}
%%%

As we have seen, for rotationally invariant QED Lorentz symmetry is \textit{extremely} well tested, with strong constraints all the way up to dimension six operators.  Lest the reader get a false impression, we also should mention that many other sectors of the standard model, from neutrons to mesons both with and without rotation invariance also have tight limits set on any possible deviation from Lorentz invariance.  

There are two areas where there remains immediate work to be done. First, there is one current caveat in regards to the dimension six operator constraints that needs to be resolved.  As we have seen, the dimension six constraints mostly rely on the physics of the GZK feature of the UHECR spectrum. More specifically, UHECR constraints rest upon the hypothesis, not in contrast with any previous experimental evidence, that protons constitute the majority of UHECRs above $10^{19}~\eV$. Recent PAO \cite{Abraham:2010yv} and Yakutsk \cite{Glushkov:2007gd} observations, however, showed hints of an increase of the average mass composition with rising energies up to $E \approx 10^{19.6}~\eV$, although still with large uncertainties mainly due to the proton-air cross-section at ultra high energies. Hence, experimental data suggests that heavy nuclei can possibly account for a substantial fraction of UHECR arriving on Earth. Furthermore the evidence for correlations between UEHCR events and their potential extragalactic sources \cite{Cronin:2007zz}--- such as active galactic nuclei (mainly blazars) --- has not improved with increasing statistics. This might be interpreted as a further hint that a relevant part of the flux at very high energies should be accounted for by heavy ions (mainly iron) which are much more deviated by the extra and inter galactic magnetic fields due to their larger charge with respect to protons (an effect partially compensated by their shorter mean free path at very high energies). If consequently one conservatively decides to momentarily suspend his/her judgment about the evidence for a GZK feature, then he/she would lose the constraints at $n=4$ on the QED sector\footnote{This is a somewhat harsh statement given that it was shown in \cite{Hooper:2010ze} that a substantial (albeit reduced) high energy gamma ray flux is still expected also in the case of mixed composition, so that in principle the previously discussed line of reasoning based on the absence of upper threshold for UHE gamma rays might still work.} as well as very much weaken the constraints on the hadronic one. 

Assuming that current hints for a heavy composition at energies $E \sim 10^{19.6}~\eV$ \cite{Abraham:2010yv} may be confirmed in the future, that some UHECR is observed up to $E \sim 10^{20}~\eV$ \cite{Abraham:2010mj}, and that the energy and momentum of the nucleus are the sum of energies and momenta of its constituents (so that the parameter in the modified dispersion relation of the nuclei is the same of the elementary nucleons) one could place a first constraint on the absence of spontaneous decay for nuclei which could not spontaneously decay without LV.\footnote{UHE nuclei suffer mainly from photodisintegration losses as they propagate in the intergalactic medium. Because photodisintegration is indeed a threshold process, it can be strongly affected by LV. According to \cite{Saveliev:2011vw}, and in the same way as for the proton case, the mean free paths of UHE nuclei are modified by LV in such a way that the final UHECR spectra after propagation can show distinctive LV features. However, a quantitative evaluation of the propagated spectra has not been performed yet.}  Such a constraint would place bounds on subluminal protons, because in this case the energy of the emitted nucleon is lowered with respect to the LI case until it ``compensates'' the binding energy of the nucleons in the initial nucleus in the energy-momentum conservation. A complementary constraint, an upper limit on superluminal protons, can be obtained from the absence of vacuum \v{C}erenkov emission.  If UHECR are mainly iron at the highest energies the constraint is given by $\zeta_p \lesssim 2\times10^{2}$ for nuclei observed at $10^{19.6}~\eV$ (and $\zeta_{p} \lesssim 4$ for $10^{20}~\eV$), while for helium it is $\zeta_{p} \lesssim 4\times10^{-3}$ \cite{Saveliev:2011vw}. 

Second, the gravitational sector of Lorentz symmetry violation is currently less constrained.  A large region of the parameter space of Einstein-aether theory remains unconstrained, while atom interferometry is starting to probe matter-tensor couplings that are sensitive to the dynamics of the Lorentz violating tensors.  Useful modified gravity theories that also evade solar system tests, such as galileons~\cite{Nicolis:2008in}, or allow for better UV behavior, such as the aforementioned Ho\v rava--Lifshitz  gravity, yield interesting Lorentz symmetry violating gravitational phenomenology.  Hence fully exploring Lorentz violation in the gravitational sector is currently an important area that requires further progress.
%
%In conclusion, we can see that the while much has been done still plenty is to be explored for dimension six operators. In particular, all of our existing constraints on the dimension six operators for QED are based on the GZK effect (either directly or indirectly) whose detection is still uncertain. It would be nice to be able to cast comparable constrains using more reliable observations, but at the moment it is unclear what reaction could play this role. Similarly, new ideas like the one of gravitational confinement \cite{Pospelov:2010mp} presented in section \ref{gravconf}, seems to call for much deeper investigation of LV phenomenology in the purely gravitational sector.   We have gone along way into exploring the possible phenomenology of Lorentz breaking physics and pushed well beyond expectations the tests of this fundamental symmetry of Nature, however still much seems to await along the path.

\begin{acknowledgement}
We wish to thank Luca Maccione for useful insights, discussions and feedback on the manuscript preparation.
\end{acknowledgement}

\bibliographystyle{spphys}	% (uses file "arnuke_revised.bst")
\bibliography{references-LC-1}

\end{document}